\documentclass[sigconf]{acmart}
\usepackage{multirow}
\usepackage{enumitem}
\usepackage{subcaption}
\usepackage[many]{tcolorbox}
\usepackage{colortbl}
\newtcolorbox[auto counter,number within=section]{observation}[2][]{
    colback=gray!5, 
    colframe=gray!35!black,
    colbacktitle=gray!35!black, 
    coltitle=white, 
    fonttitle=\bfseries, 
    title=Observation\ #2,
    enhanced,
    attach boxed title to top left={yshift=-2mm, xshift=0.5cm}, 
    #1
}
\usepackage{threeparttable}
\usepackage{enumitem}
\usepackage{algorithm}
\usepackage{algpseudocode}
\usepackage{graphicx}
\usepackage{lipsum}
\usepackage{amsmath}
\usepackage{xcolor}
\setlist[itemize]{left=0pt}

\AtBeginDocument{%
  \providecommand\BibTeX{{%
    \normalfont B\kern-0.5em{\scshape i\kern-0.25em b}\kern-0.8em\TeX}}}

\begin{document}

\title[InverTune: Removing Backdoors from Multimodal Contrastive Learning Models]{InverTune: Removing Backdoors from Multimodal Contrastive Learning Models via Trigger Inversion and Activation Tuning}

\author{Mengyuan Sun$^{1,*}$, Yu Li$^{1,*}$, Yuchen Liu$^{1}$, Bo Du$^{2}$, Yunjie Ge$^{3,\dagger}$}
\affiliation{%
\institution{$^1$School of Cyber Science and Engineering, Wuhan University \\ \hspace{0.3em} $^2$School of Computer Science, Wuhan University \hspace{0.3em} $^3$Institute for Math \& AI, Wuhan University}
\city{}
\country{}}
\affiliation{%
\institution{}
\city{}
\country{}}
\thanks{$^{*}$ The first two authors contributed equally to this work. 

$^{\dagger}$Corresponding author.}

\begin{abstract}

Multimodal contrastive learning models like CLIP have demonstrated remarkable vision-language alignment capabilities, yet their vulnerability to backdoor attacks poses critical security risks. Attackers can implant latent triggers that persist through downstream tasks, enabling malicious control of model behavior upon trigger presentation. Despite great success in recent defense mechanisms, they remain impractical due to strong assumptions about attacker knowledge or excessive clean data requirements. In this paper, we introduce InverTune, the first backdoor defense framework for multimodal models under minimal attacker assumptions, requiring neither prior knowledge of attack targets nor access to the poisoned dataset. Unlike existing defense methods that rely on the same dataset used in the poisoning stage, InverTune effectively identifies and removes backdoor artifacts through three key components, achieving robust protection against backdoor attacks. Specifically, InverTune first exposes attack signatures through adversarial simulation, probabilistically identifying the target label by analyzing model response patterns. Building on this, we develop a gradient inversion technique to reconstruct latent triggers through activation pattern analysis. Finally, a clustering-guided fine-tuning strategy is employed to erase the backdoor function with only a small amount of arbitrary clean data, while preserving the original model capabilities. 
Experimental results show that InverTune reduces the average attack success rate (ASR) by 97.87\% against the state-of-the-art (SOTA) attacks while limiting clean accuracy (CA) degradation to just 3.07\%. 
This work establishes a new paradigm for securing multimodal systems, advancing security in foundation model deployment without compromising performance.
\end{abstract}


\renewcommand\footnotetextcopyrightpermission[1]{}
\setcopyright{none}
\maketitle

\section{Introduction}

Multimodal contrastive learning (MCL) has revolutionized vision-language alignment, enabling breakthroughs in various challenging tasks like zero-shot classification~\cite{larochelle2008zero,ye2017zero,christensen2023image,qian2024online}, image captioning~\cite{vinyals2015show,mokady2021clipcap,barraco2022unreasonable,cho2022fine}, and visual question answering~\cite{antol2015vqa,eslami2021does,eslami2023pubmedclip}. Models like CLIP~\cite{radford2021learning} align images and text into a shared embedding space through web-scale pretraining, achieving remarkable generalization without task-specific fine-tuning. Subsequent advancements, including ALIGN~\cite{jia2021scaling} and CoOp~\cite{zhou2022learning}, further enhance MCL’s robustness, cementing its role in modern multimodal systems.

While MCL models have achieved impressive success in various tasks, they are not without vulnerabilities. Especially, the reliance on large-scale, web-crawled training data exposes MCL models to backdoor attacks, in which adversaries implant hidden triggers to manipulate downstream task behavior. Different from unimodal attacks, backdoor attacks against MCL exploit target cross-modal alignment mechanisms, inducing misalignment between visual and textual representations. For example, BadCLIP~\cite{liang2024badclip} poisons training data to associate a visual trigger with mismatched text labels. 
When users unknowingly fine-tune their models with these poisoned data under downstream tasks, a backdoor can be stealthily embedded into the model, enabling adversaries to manipulate deployed systems. Owing to the widespread practice of fine-tuning untrusted pre-trained models, these vulnerabilities are further exacerbated, creating an urgent need for defenses against backdoors.

Recently, many approaches have been proposed to detect or purify backdoors. Detection methods~\cite{feng2023detecting} can only identify poisoned encoders but do not provide remediation. Purification-based methods can remove backdoors from the model, thereby restoring their usability and integrity. Yet, they either require impractical amounts of clean data~\cite{bansal2023cleanclip}, need precise hyperparameter tuning~\cite{zhang2024defending}, or cause a terrible trade-off between model performance and defensive effectiveness~\cite{DBLP:journals/corr/abs-2409-17601,kuang2024adversarial}. 
These shortcomings raise a critical question: \textit{Can we develop a practical defense that simultaneously eliminates backdoors and preserves clean-task performance under reasonable assumptions?}

Addressing this question is particularly challenging in the context of MCL models. In conventional single-modal classification models, defenders can effectively identify backdoor targets through exhaustive testing across a predefined discrete label space, enabling precise backdoor mitigation. 
However, the open-vocabulary nature of MCL~\cite{chen2023exploring} fundamentally invalidates such enumeration-based approaches. Moreover, in the single-modal classification model, only the target label is affected so that the defender can use this particularity to identify the backdoor information. In an MCL model, the impact of a backdoor may not be limited to a single label, such as multiple tokens, making it difficult for defenders to distinguish. 
Crucially, if the target labels in MCL models can be accurately identified, it would greatly simplify the process of backdoor mitigation.

\begin{figure*}[t!] 
  \centering
  \includegraphics[width=.95\linewidth]{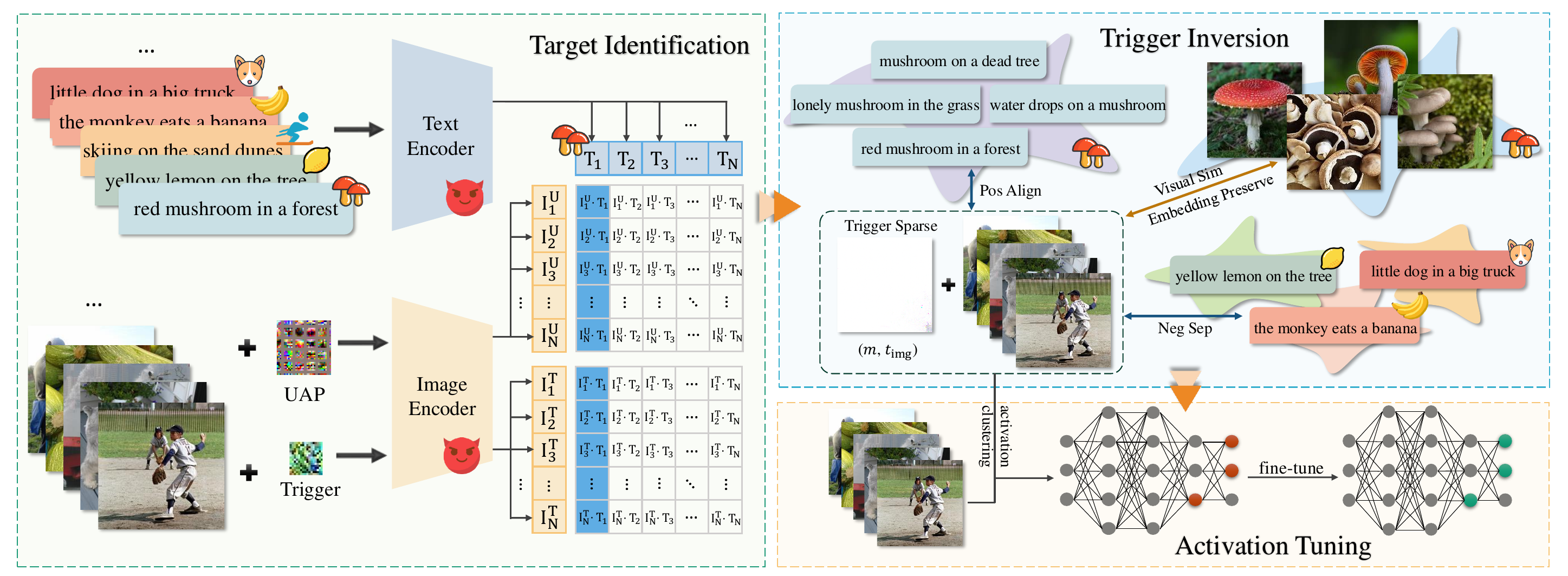}
  \caption{Target label identification, trigger inversion, and activation tuning: InverTune’s framework for backdoor removal (illustrated with a mushroom-targeted example).}
  \label{fig:framework}
\end{figure*}

In response to the above question, we propose InverTune, a new backdoor defense framework for MCL models, which could remove backdoors while preserving model performance under manageable assumptions. The workflow of InverTune is to first identify the backdoor information and then purify it correctly. 
First, InverTune identifies target labels by exploiting a key observation: backdoored models exhibit unique vulnerabilities to adversarial perturbations compared to clean models. This approach provides precise target label detection and significantly reduces computational overhead to exhaustive search. 
Second, we propose a dual-space optimization strategy that jointly analyzes the visual embedding space and cross-modal alignment space. By minimizing the discrepancy between perturbed and target embeddings across both spaces, InverTune accurately isolates trigger signatures while preserving the model’s original feature representations. Third, we perform clustering-based fine-tuning to selectively recalibrate these neurons corresponding to backdoors, suppressing malicious functionality without sacrificing clean-task performance. 
We evaluate the effectiveness of InverTune against six MCL backdoor attacks, including the SOTA attack BadCLIP, and compare it with four leading defense approaches. The results show that, on both ImageNet classification and MSCOCO image-to-text retrieval tasks, we reduce most attack success rate (ASR) to within 1.0\%, achieving an average ASR decrease of 89.88\% and 97.58\% respectively, demonstrating SOTA defense performance. Meanwhile, we maximally preserve the model utility, achieving an average clean accuracy (CA) of 54.96\% and 69.47\%. This demonstrates a good balance between backdoor removal and model utility preservation during the defense process.

Our contributions can be summarized as follows:

\begin{itemize}
    \item To the best of our knowledge, We are the first to identify backdoor target labels in MCL models. This discovery not only enables backdoor risk verification but also unlocks precise, low-cost defense mechanisms by directly identifying the root of attacks. 
     
    \item We introduce InverTune, a novel three-step defense framework that integrates backdoor label identification, gradient-guided trigger inversion, and activation-aware fine-tuning, requiring only reasonable amounts of data. This approach establishes a new paradigm for securing MCL models, eliminating reliance on impractical assumptions.
       
    \item Extensive experimental results show that InverTune has strong defense power. Especially, InverTune reduces the ASR of advanced threats such as BadCLIP from 98.36\% to 0.49\%, outperforming existing defenses by 17.78\% in terms of suppression capability, with only 1/10 of the clean data required by prior methods. Notably, it achieves an average Top-10 CA of 69.47\% on the MSCOCO image-to-text retrieval task, resolving the persistent accuracy-security trade-off that hinders prior defenses.
\end{itemize}

\begin{figure}[t]
  \centering
  \begin{minipage}[b]{0.45\linewidth}
    \centering
    \includegraphics[width=\linewidth]{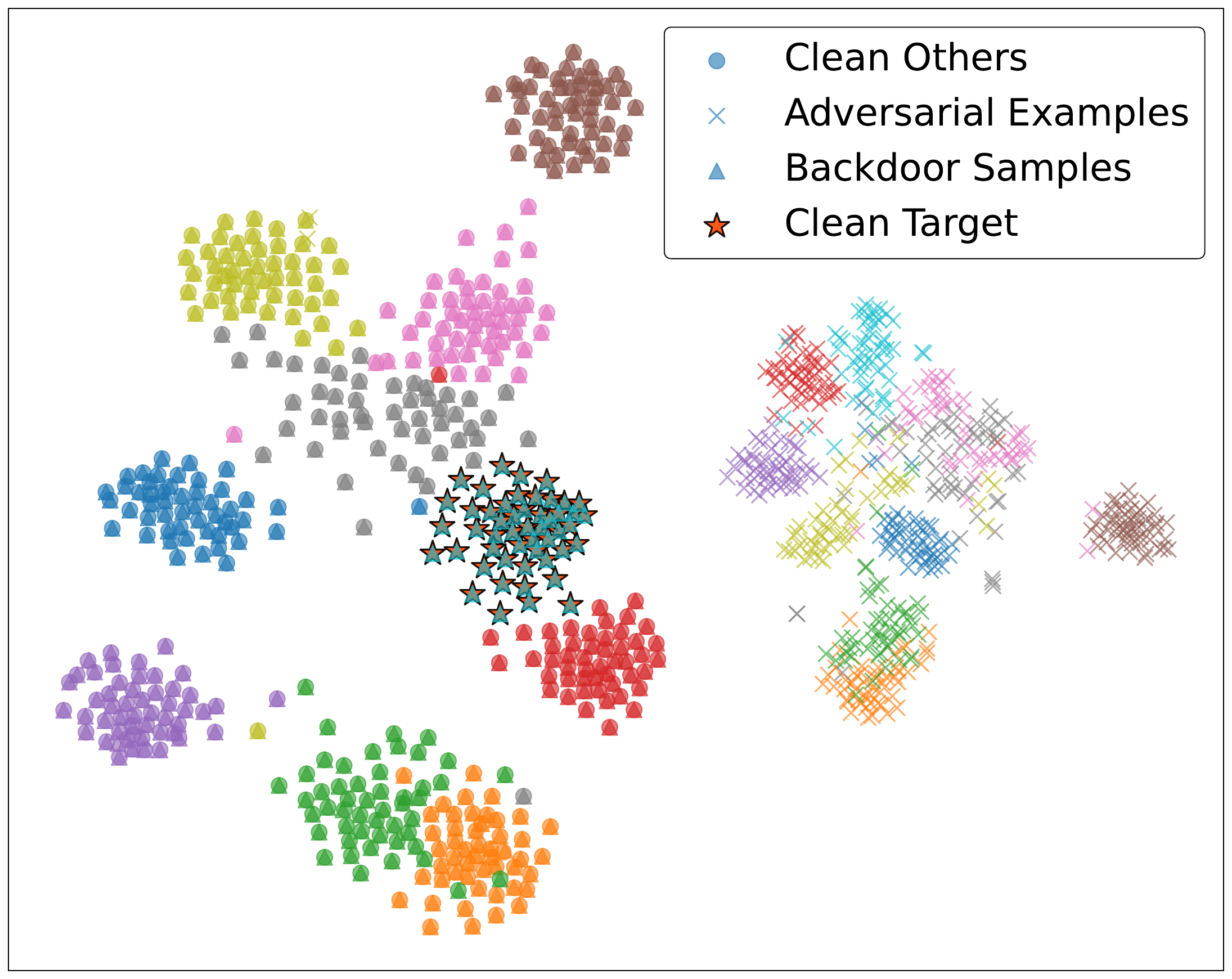}
    \caption*{(a) Clean encoder.}
  \end{minipage}
  \hfill
  \begin{minipage}[b]{0.45\linewidth}
    \centering
    \includegraphics[width=\linewidth]{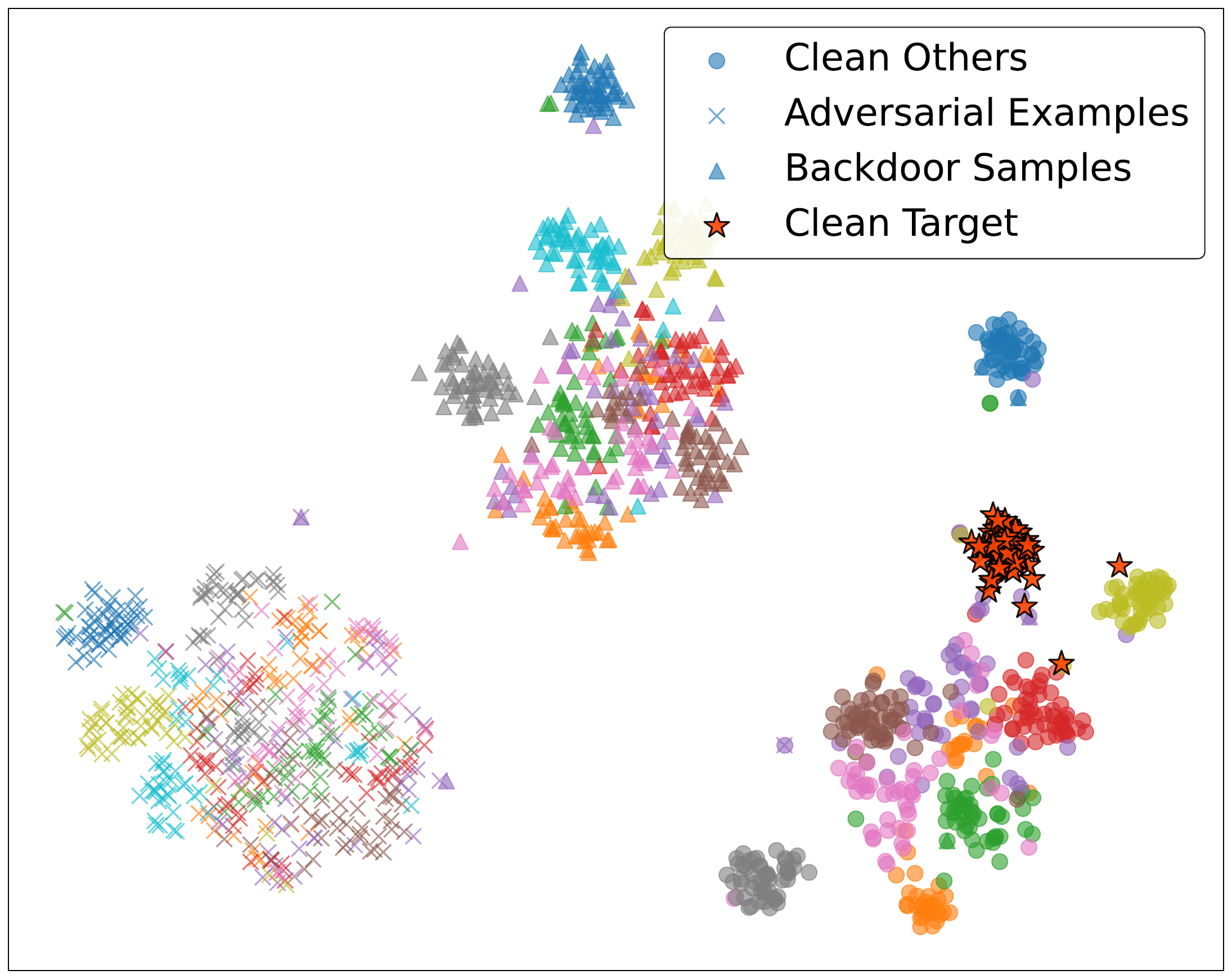}
    \caption*{(b) Poisoned encoder.}
  \end{minipage}
  \vspace{-0.8em}
  \caption{The t-SNE plots of clean samples, backdoor samples, and adversarial examples in (a) the clean model and (b) the backdoored model.
  }
  \label{fig:tsne_features}
\end{figure}

\section{Threat Model}

\textbf{Attacker.}
We follow the SOTA settings \cite{liang2024badclip} for backdoor attacks in MCL models, specifically targeting the vision encoder. We assume that the attacker can construct a poisoned fine-tuning dataset and knows the model architecture and parameters. The attacker’s goal is to implant a backdoor into the pre-trained CLIP model such that the model behaves normally on benign inputs but outputs incorrect results when exposed to inputs with triggers. To achieve this, the attacker injects a small portion of poisoned samples into the fine-tuning dataset, introducing visual triggers. The attacker then fine-tunes the pre-trained model using this poisoned dataset, manipulating the model’s responses to visual triggers. Once the vision encoder is backdoored, the attacker has no control over downstream applications or tasks using the model.

\noindent\textbf{Defender.}
To conduct a practical defense, we assume that the defender has no access to the pretraining dataset or the poisoned fine-tuning dataset, and is unaware of the backdoor attack's target. Furthermore, the defender either has no access to the full clean dataset or only possesses a limited amount of clean data. The primary goal of the defender is to neutralize the backdoors while maintaining the model's original performance on the clean data.

\section{InverTune: Detailed Construction}

As illustrated in Figure~\ref{fig:framework}, our proposed method, InverTune, mitigates backdoor attacks in MCL models through a three-step process: adversarial perturbation-based target identification, trigger inversion, and activation clustering-based fine-tuning.

\subsection{Target Identification}
\label{sec:step1}

\begin{figure}[t]
  \centering
  \includegraphics[width=0.8\linewidth, height=0.51\linewidth]{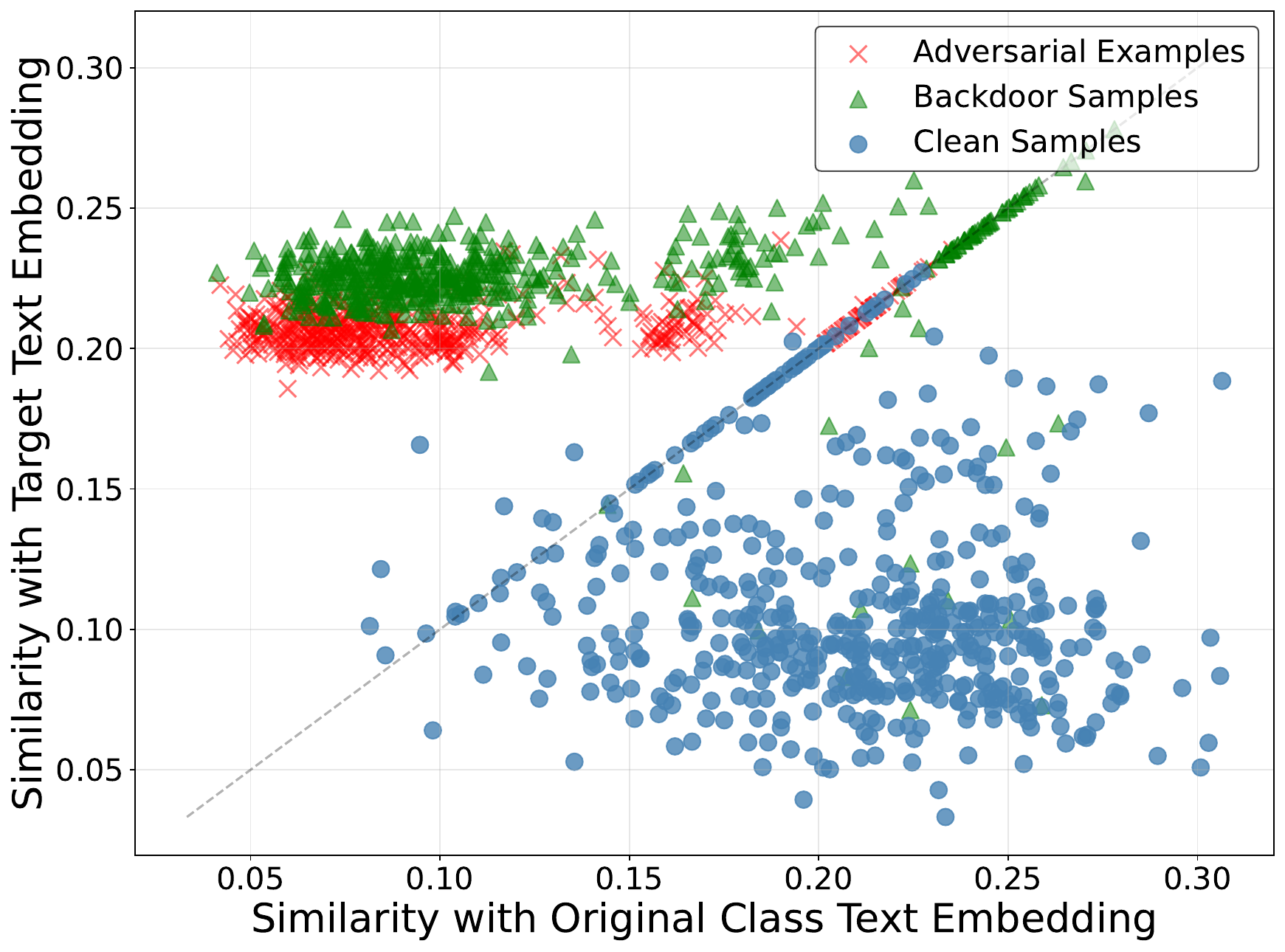}
  \caption{Image-text similarity shift: backdoor and adversarial examples are closer to target text than to original text.}
  \label{fig:similarity_comparison}
\end{figure}

Recent studies~\cite{mu2023progressive, niu2024towards} reveal that backdoored models exhibit distinct characteristics in feature representation and vulnerability within target classes. Single-modal backdoored models establish strong associations between target class labels and both robust features and backdoor features. Hence, normal and backdoored samples of the target class cluster closely in the latent space. Besides, untargeted adversarial attacks would inadvertently exploit backdoor pathways so that the optimization process for generating adversarial perturbations leans to converge toward backdoor triggers, causing attack outcomes to disproportionately favor the backdoor target label. In contrast, benign models exhibit approximately uniform label distribution for adversarial examples. This phenomenon has motivated backdoor defense strategies that leverage adversarial example analysis for trigger inversion.  In the MCL domain, Kuang et al.~\cite{kuang2024adversarial} directly utilize the insight to optimize universal adversarial perturbations followed by anti-learning purification. However, their defense performance is unsatisfactory.

Inspired by the above finding and result,  we try to understand how the backdoor affects the target class in the MCL model.  To achieve this, we take a SOTA MCL backdoor attack method, BadCLIP, as an example.
Specifically, we visualize the visual encoder features of backdoor samples, adversarial examples generated using AdvCLIP~\cite{zhou2023advclip}, and clean images from 10 randomly selected categories, including the target category. Based on Figure~\ref{fig:tsne_features}, we find new observations different from those in the single-modal model.

\begin{figure}[t]
  \centering
  \begin{minipage}[b]{0.48\linewidth}
    \centering
    \includegraphics[width=\linewidth]{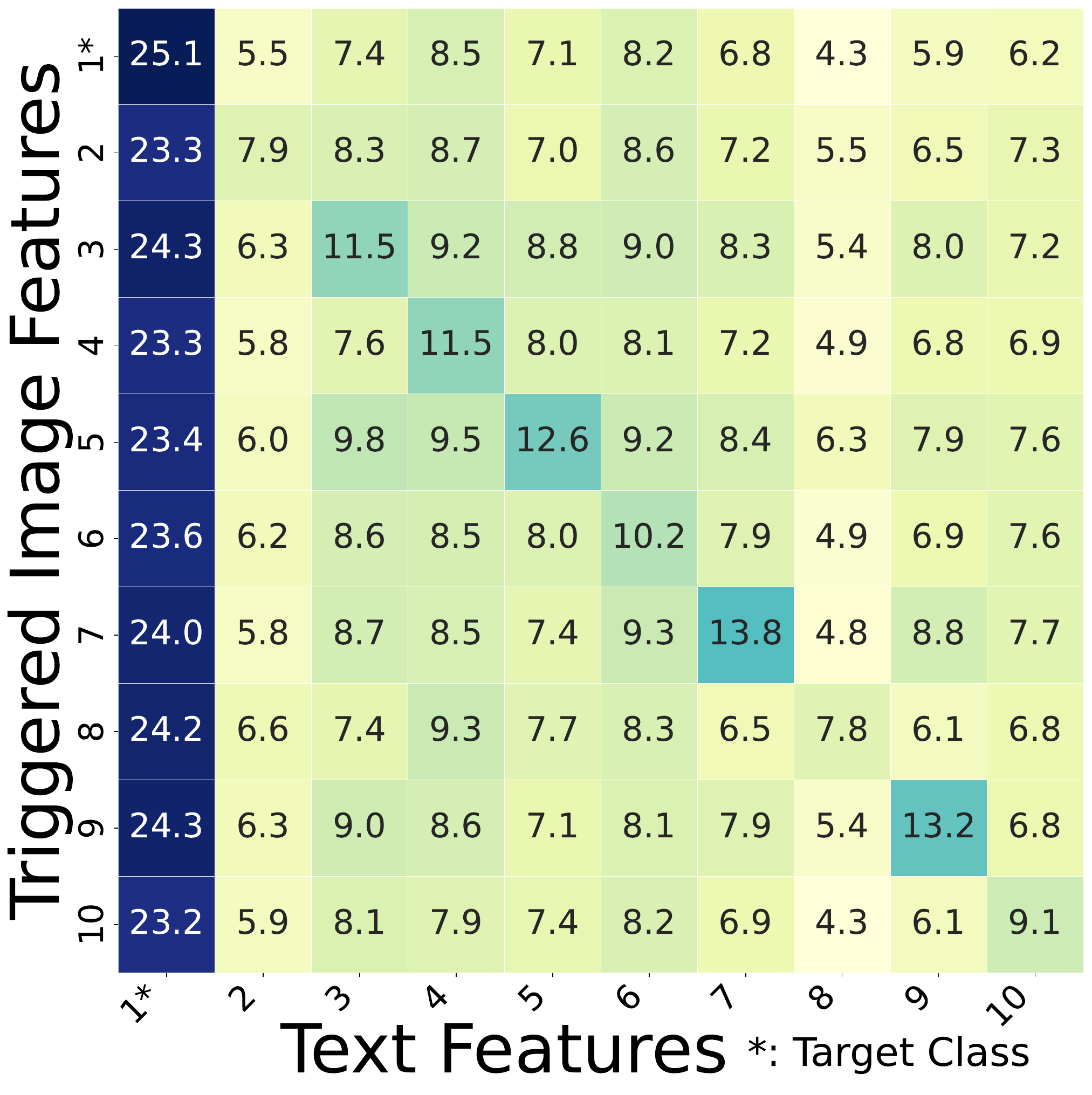}
    \vspace{-2em}
    \caption*{(a) Backdoor samples.}
  \end{minipage}
  \hfill
  \begin{minipage}[b]{0.48\linewidth}
    \centering
    \includegraphics[width=\linewidth]{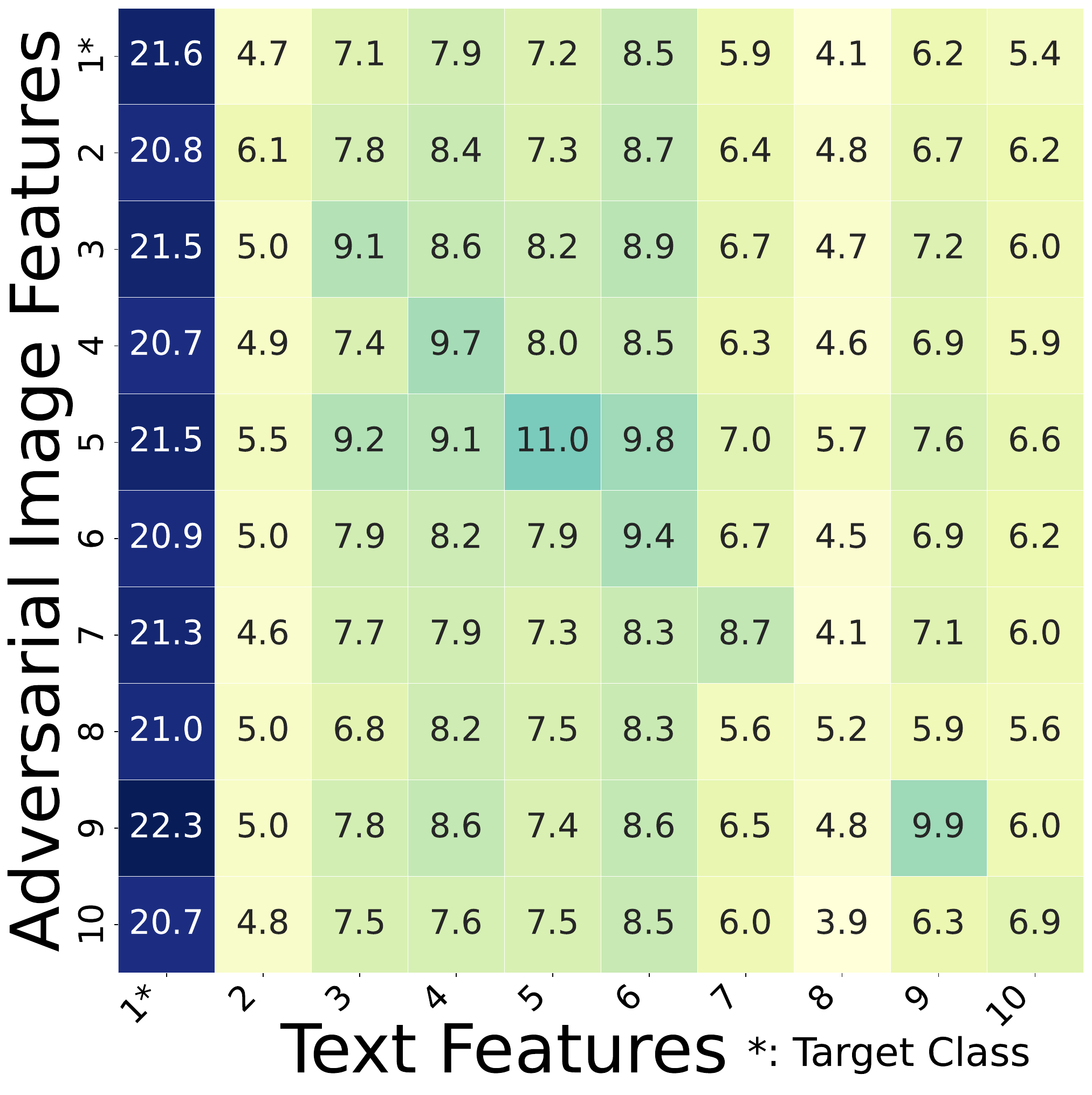}
    \vspace{-2em}
    \caption*{(b) Adversarial examples.}
  \end{minipage}
  \vspace{-1em}
  \caption{Similarity matrices between image and text features under different attack scenarios.}
  \label{fig:similarity_matrices}
\end{figure}        

\begin{observation}{I}
  Backdoor samples form distinct clusters rather than merging with target class features.
  \end{observation}
  
We find that although BadCLIP's dual-embedding optimization reduces the visual embedding distance between poisoned samples and target class samples, samples with triggers form a new cluster in visual features and do not become closer to the target class samples. Moreover, by observing adversarial examples, we also find similar results. To understand it more, we calculate the similarity between backdoor samples and adversarial examples, as shown in Figure~\ref{fig:similarity_comparison} and Figure~\ref{fig:similarity_matrices}. Based on all results, we notice another observation.

\begin{observation}{II}
  Adversarial attacks tend to exploit backdoor-induced weaknesses rather than direct trigger mimicry.
\end{observation}

Since adversarial examples and backdoor samples remain significantly distant in terms of feature space, this suggests that adversarial examples do not directly mimic the features of backdoor samples. Based on Figure~\ref{fig:similarity_matrices}, we can find that most adversarial examples have higher similarity with the text features of the target class, showing the backdoor also affects the adversarial attacks. This suggests backdoors reconfigure multimodal decision boundaries, creating ``vulnerability zones'' that adversarial attacks preferentially exploit. As a result, adversarial attacks are more likely to exploit this vulnerability, causing higher confusion and increasing the chances of misclassification into the target class.

\noindent\textbf{Identification Strategy.} Building on these insights, we develop a target label identification strategy through differential analysis of adversarial misclassification patterns. 
Specifically, given a suspected compromised model, we construct a universal adversarial perturbation designed to induce systematic misclassification across all input images. 
We then compare the model's output distribution on adversarially perturbed samples $P_{\text{adv}}(y)$ against its predictions on clean samples $P_{\text{clean}}(y)$. The target label $y_t$ is identified as the class exhibiting the maximum increase in prediction frequency:
\begin{equation}
    y_t = \arg \max\limits_{y \in \mathcal{Y}}{(P_{\text{adv}}(y)-P_{\text{clean}}(y))}.
\end{equation}
This differential analysis isolates attack-induced bias from natural model tendencies, leveraging the intrinsic concentration property of backdoor attacks: backdoored models consistently steer misclassified samples toward the target label with disproportionate frequency. The identified target label then serves as the foundation for subsequent backdoor mitigation through gradient-guided trigger inversion and activation suppression.

\subsection{Trigger Inversion}
\label{sec:step2}

Unlike traditional single-modal backdoor attacks where the target is a specific class label, multimodal backdoor attacks in CLIP exploit the complex cross-modal alignment between visual and textual representations. This fundamental difference requires a specialized approach to trigger inversion that addresses the unique characteristics of multimodal contrastive learning models.

\noindent\textbf{Multimodal Trigger Inversion Challenges.}
Traditional backdoor inversion methods~\cite{wang2019neural, guo2019tabor, wang2023unicorn} designed for classification models cannot be directly applied to multimodal models like CLIP for several key reasons. (1) In CLIP, backdoor attacks operate by creating malicious alignments between visual triggers and textual targets across modalities. This cross-modal interaction is fundamentally different from the class boundary manipulation in traditional classification models, as it requires simultaneous optimization over both image and text embeddings. (2) CLIP projects both images and text into a shared high-dimensional embedding space, where the backdoor behavior is determined by the alignment between these modalities. This shared space introduces additional complexity compared to the discrete class labels used in traditional models, as the backdoor functionality depends on the relative positions of embeddings rather than direct class mappings.  (3) CLIP's zero-shot capabilities~\cite{zhou2023zegclip} allow it to generalize to unseen classes and concepts, which backdoors can exploit in ways that are not observable in traditional models. This makes it challenging to detect and invert triggers, as the backdoor behavior may manifest differently across various downstream tasks.

\noindent\textbf{Dual-Space Trigger Optimization.}
To address these challenges, we propose a novel dual-space trigger inversion approach that explicitly considers both the visual embedding space and the cross-modal alignment.
Specifically, given a clean input image $x$, we parameterize the trigger as a mask-pattern pair $ (m, t_{\text{img}})$, where the backdoor sample  $\tilde{x}$  is generated via element-wise composition:
\begin{equation}
  \tilde{x} = m \odot t_\text{img} + (1-m) \odot x,
\end{equation}
where $m$ denotes the mask, $t_\text{img}$ represents the trigger pattern, and $\odot$ denotes element-wise multiplication. 
Our framework integrates four synergistic loss components to ensure precise trigger reconstruction while preserving stealthiness: Cross-Modal Alignment, Embedding Space Preservation, Visual Similarity, and Trigger Sparsity.
Detailedly, Cross-Modal Alignment is formulated using the InfoNCE~\cite{oord2018representation} loss to force the visual trigger embeddings to align with the identified target text  $y_t$ while diverging from non-target classes. The contrastive loss can be expressed as: 
 \begin{equation}
      \mathcal{L}_\text{align} = -\log\frac{\exp(\text{sim}(E_I(\tilde{x}), E_T(y_t))/\tau)}{\sum_{j=1}^{N}\exp(\text{sim}(E_I(\tilde{x}), E_T(y_j))/\tau)} ,  
\end{equation}
where $E_I$ and $E_T$ are the image and text encoders of the suspected model, $y_j$ iterates over all class prompts including the target, $\tau$ is the temperature parameter controlling the sharpness of the distribution, and $N$ is the number of considered classes. 
Then we employ the embedding space preservation loss to prevent backdoor samples from excessively shifting toward the target class's textual embedding, thereby preserving the embedding structure and maintaining a stable data distribution to safeguard generalization. It is formulated as follows:
\begin{equation}
      \mathcal{L}_\text{emb} = D(\frac{E_I(\tilde{x})}{\|E_I(\tilde{x})\|_2}, \frac{E_I(x)}{\|E_I(x)\|_2}),
    \end{equation}
where $D(\cdot)$ means a distance function. Here we employ the widely-used $L_2$-norm distance metric. 
Considering the attacker's goal, where the backdoor sample must remain visually similar to the original, we introduce a visual similarity loss as follows:
\begin{equation}
      \mathcal{L}_\text{sim} = 1 - \text{SSIM}(\tilde{x}, x),
      \label{eqsim}
    \end{equation}
where $\text{SSIM}(\cdot)$ function computes the structural similarity between two given images~\cite{wang2004image}. 
Although the loss function $\mathcal{L}_\text{sim}$ can make the backdoor sample as similar as possible to the original sample, it does not ensure the imperceptibility of the backdoor trigger. Therefore, we introduce the trigger sparsity loss to further constrain the trigger as follows:
\begin{equation}
      \mathcal{L}_\text{mask} = \|m\|_1.
    \end{equation} 
To obtain the trigger pattern and mask, we optimize the four loss functions concurrently. Therefore, the total loss can be written as the weighted combination of these objectives:
\begin{equation}
  \label{eq:inver_loss}
  \mathcal{L}_\text{inver} = \lambda_1\mathcal{L}_\text{align} + \lambda_2\mathcal{L}_\text{emb} + \lambda_3\mathcal{L}_\text{sim} + \lambda_4\mathcal{L}_\text{mask},
\end{equation}
where $\lambda_1$, $\lambda_2$, $\lambda_3$, and $\lambda_4$ are weighting coefficients for each term.

\subsection{Activation Tuning}

Building upon the inverted trigger obtained in Section \ref{sec:step2}, we propose an activation-based fine-tuning strategy specifically tailored for multimodal contrastive learning models like CLIP. This approach leverages the unique activation patterns induced by backdoor triggers in the shared embedding space of multimodal models.

\noindent\textbf{Key Insight.}  
Backdoor triggers in the MCL model exploit the cross-modal alignment mechanism, creating distinct activation signatures in specific layers. By identifying and selectively fine-tuning these critical neurons, we can effectively neutralize the backdoor while preserving the model's multimodal capabilities.

\noindent\textbf{Layer Selection.}  

Inspired by prior findings~\cite{10179375, liu2019abs} in CNN architectures where backdoor patterns predominantly affect deeper network layers, we first identify the most responsive layers to backdoor activation in MCL models. For each layer, we quantify backdoor sensitivity through normalized activation divergence:
  \begin{equation}
    \text{diff} = \frac{\| \mu_{\text{clean}} - \mu_{\text{triggered}} \|_2}{\| \mu_{\text{clean}} \|_2},
  \end{equation}
where $\mu_{\text{clean}}$ and $\mu_{\text{triggered}}$ represent the average activations of clean and triggered inputs, respectively.  Then, we compute the mean and standard deviation of activation differences across all layers. The layers with activation differences exceeding the mean by more than one standard deviation will be treated as backdoor-related. Within these critical layers, we further analyze individual neuron activation variances. Note that identifying only the neurons in the backdoor-related layer greatly reduces the time and resource overhead compared to identifying all neurons once.

\noindent\textbf{Critical Neuron Identification.}
We identify critical neurons by first measuring the impact of the trigger on layer activations. For each selected layer, we calculate the mean activation difference between the clean and trigger-affected inputs. Then, we apply K-means clustering~\cite{macqueen1967some} on the activation differences to group neurons with similar response patterns. Clustering helps address the potential variability in neuron responses. Instead of simply selecting the neurons with the largest activation difference, K-means clustering groups neurons with similar response patterns, ensuring that the neurons we capture share a common sensitivity to the backdoor.

\noindent\textbf{Fine-Tuning Process.}
Following neuron identification, we implement targeted fine-tuning to eliminate backdoor functionality while preserving clean-task performance.  Specifically, we introduce an activation alignment loss to force backdoor-sensitive neurons to exhibit similar activation patterns for clean and triggered samples:
\begin{equation}
\mathcal{L}_{\text{activation}} = \sum_{i \in \text{critical}} \|\mathbf{a}_{\text{clean}}^i - \mathbf{a}_{\text{triggered}}^i\|_2^2.
\end{equation}
This suppresses backdoor-triggered activation spikes. Moreover, to maintain original vision-language alignment capability, we introduce a cross-modal consistency loss.
\begin{equation}
    \mathcal{L}_{\text{preserve}} = \|\text{sim}(E_I(x), E_T(y)) - \text{sim}(E_I^\text{orig}(x), E_T(y))\|_2^2,
    \label{e3}
    \end{equation}
where $E_{I}^{\text{orig}}$ represents the original backdoored encoders prior to fine-tuning. This function forces the fine-tuned model to have similar normal functions to the original model.
In order to achieve both purposes, the composite optimization objective becomes:
\begin{equation}
\label{eq:finetune_loss}
\mathcal{L}_\text{tune} = \mathcal{L}_\text{activation} + \beta \mathcal{L}_{\text{preserve}},
\end{equation}
where $\beta$  is to balance the two objectives. Note that, we apply neuron masks during gradient updates to restrict fine-tuning to critical neurons. This targeted fine-tuning minimizes disruption to the model's overall performance while effectively mitigating the backdoor.

\begin{table*}[t]
  \caption{The defensive performance of InverTune across various tasks and adversarial attacks. The optimal ASR and CA values are highlighted in bold, while the second-best results are indicated with underlining.}
  \label{tab:main_1}
  \centering
  \begin{tabular}{@{}rc|cccccccccccc@{}}
    \toprule
    \multirow{2}{*}{} & \multirow{2}{*}{\textbf{Methods}} & \multicolumn{2}{c}{\textbf{BadNet}} & \multicolumn{2}{c}{\textbf{Blended}}  & \multicolumn{2}{c}{\textbf{SIG}} & \multicolumn{2}{c}{\textbf{WaNet}} & \multicolumn{2}{c}{\textbf{BadEncoder}}  & \multicolumn{2}{c}{\textbf{BadCLIP}} \\
    \cmidrule(lr){3-4} \cmidrule(lr){5-6} \cmidrule(lr){7-8} \cmidrule(lr){9-10} \cmidrule(lr){11-12}   \cmidrule(lr){13-14}  
    &  & CA $\uparrow$ & ASR $\downarrow$ & CA $\uparrow$ & ASR $\downarrow$ & CA $\uparrow$ & ASR $\downarrow$ & CA $\uparrow$ & ASR $\downarrow$ & CA $\uparrow$ & ASR $\downarrow$ & CA $\uparrow$ & ASR $\downarrow$ \\
    \midrule
    \multirow{6}{*}{\rotatebox{90}{\textbf{ImageNet}}}  
    & No Defense & 58.21 & 87.73 & 58.74 & 96.35 & 58.30 & 82.57 & 58.64 & 96.18 & 53.10 & 80.13 & 58.32 & 98.36 \\
    & FT & \underline{54.13} & 33.67 & \textbf{54.64} & 64.10 & \textbf{54.36} & 55.59 & \underline{54.59} & 58.38 & \textbf{55.98} & 19.71 & 54.16 & 86.03 \\
    & CleanCLIP & 51.92 & 4.62 & 51.38 & 52.36 & 51.42 & 36.72 & 51.45 & 24.98 & 55.29 & 5.21 & \underline{54.18} & 75.17 \\
    & CleanerCLIP & 51.91 & \underline{3.87} & 52.36 & 11.38 & 52.56 & \underline{9.89} & 51.57 & 10.94 & 52.11 & \textbf{0.19} & 51.74 & 21.16 \\
    & PAR & 53.57 & 6.03 & \underline{54.18} & \underline{0.16} & 51.96 & 22.94 & 53.89 & \underline{4.51} & 54.25 & 2.27 & 50.95 & \underline{17.78} \\
    & InverTune (Ours) & \textbf{56.12} & \textbf{0.02} & 53.50 & \textbf{0.14} & \underline{54.27} & \textbf{0.28} & \textbf{54.76}& \textbf{0.09} & \underline{55.84} & \underline{1.02} & \textbf{55.25} & \textbf{0.49} \\ 
    \midrule
    \multirow{6}{*}{\rotatebox{90}{\textbf{MSCOCO}}} 
    & No Defense & 69.94 & 95.88 & 71.20 & 99.76 & 70.28 & 97.42 & 71.16 & 99.60 & 72.07 & 98.13 & 71.32 & 99.28\\
    & FT & \underline{68.83} & 39.09 & \textbf{69.53} & 67.51 & \underline{68.92} & 63.67 & \textbf{69.70} & 70.41 & \textbf{68.77} & 25.47 & \underline{68.25} & 88.54\\
    & CleanCLIP & 65.03 & 14.17 & 63.70 & 55.47 & 64.09 & 38.71 & 67.61 & 64.83 & 67.56 & 13.42 & 66.53  & 84.55\\
    & CleanerCLIP & 65.73 & \underline{7.94} & 68.82 & 14.93 & 65.98 & \underline{14.31} & 64.67 & 15.01 & 66.39 & \underline{3.41} & 65.21 & 30.41  \\
    & PAR & 68.42 & 15.43 & 68.11 & \textbf{0.37} & 66.64 & 31.09 & 68.28 & \underline{7.83} & 67.42 & 4.30 & 65.73 & \underline{16.47} \\
    & InverTune (Ours) & \textbf{71.12} & \textbf{0.04}& \underline{69.16} & \underline{0.52} & \textbf{69.94} & \textbf{1.12} & \underline{68.98} & \textbf{0.48} & \underline{68.02} &\textbf{1.73} &\textbf{ 69.58 }& \textbf{0.68} \\  
    \bottomrule 
  \end{tabular}
\end{table*}

\section{Experiment}

\subsection{Experiment Setup}

\textbf{Models.} We adopt OpenAI's open-source CLIP model~\cite{radford2021learning} as our pretrained base, using RN50 as the default backbone architecture. For a comprehensive evaluation, we extend our analysis to RN101, ViT-B/16, and ViT-B/32 architectures in Section \ref{sec:targetlabel}.

\noindent\textbf{Datasets.} Following~\cite{liang2024badclip}, we use a 500K subset of CC3M~\cite{sharma2018conceptual} for poisoning the clean CLIP model. The evaluation framework covers two key tasks: zero-shot classification on ImageNet-1K validation set~\cite{ILSVRC15} and image-to-text retrieval on Microsoft COCO 2017~\cite{lin2014microsoft}.

\noindent\textbf{Backdoor Attacks.} We evaluate our defense method against four representative single-modal backdoor attack methods: BadNet~\cite{gu2017badnets}, Blended~\cite{chen2017targeted}, SIG~\cite{barni2019new}, and WaNet~\cite{nguyen2021wanet}. Additionally, we include one self-supervised learning backdoor attack on a pretrained encoder, BadEncoder~\cite{jia2022badencoder}, and the SOTA CLIP-specific backdoor attack, BadCLIP~\cite{liang2024badclip}.
We randomly select ``mushroom'' as the target label. Experiments with other target labels are presented in Section ~\ref{sec:targetlabel}. Following the settings of \cite{liang2024badclip}, we set the poisoning rate to 0.3\%. 

\noindent\textbf{Implementation Details.} For the InverTune, we set $\lambda_1 = 5.0$, $\lambda_2 = 0.5$, $\lambda_3 = 1.0$, and $\lambda_4 = 0.01$ for the trigger inversion loss in Equation~\eqref{eq:inver_loss}. The optimization is performed using the Adam optimizer with a learning rate of $1 \times 10^{-2}$. For activation tuning, we set $\beta = 0.5$ for the fine-tuning loss in Equation~\eqref{eq:finetune_loss}, use a learning rate of $8 \times 10^{-6}$, and train for 200 epochs. In terms of data usage, InverTune employs a 50K subset of the ImageNet-1K training set~\cite{ILSVRC15}, which is only 1/10 the size of the data used by other baselines. In the activation tuning step, we require only a single batch (predefined as 64) of arbitrary clean data. All experiments are conducted on an NVIDIA A100 GPU. More details are provided in our Supplementary Materials.

\noindent\textbf{Baselines.} We compare our method against several advanced backdoor defense techniques, including CleanCLIP~\cite{bansal2023cleanclip}, CleanerCLIP~\cite{DBLP:journals/corr/abs-2409-17601}, PAR~\cite{singh2024perturb}, as well as Fine-Tuning (FT)~\cite{bansal2023cleanclip} as the baselines.

\noindent\textbf{Evaluation Metrics.}
We evaluate the effectiveness of our method using the following metrics. \textit{Clean Accuracy} (CA): For zero-shot classification tasks, CA quantifies the model's Top-1 prediction accuracy on unperturbed inputs. For image-to-text retrieval scenarios, it measures the proportion of clean queries successfully matching ground-truth captions within the Top-10 retrieved results. Higher CA values indicate better preservation of the model's normal capabilities. \textit{Attack Success Rate} (ASR):  For classification, ASR represents the percentage of triggered samples misclassified to target labels. For image-to-text retrieval tasks, ASR is the percentage of triggered inputs that retrieve target-related text in the Top-10 results. Lower ASR scores demonstrate superior backdoor mitigation.

\subsection{InverTune Performance}
\label{sec:performance}

\noindent\textbf{Defensive Performance}.
The experimental results in Table~\ref{tab:main_1} show InverTune's superior defensive capabilities across multiple attack scenarios. Our method achieves state-of-the-art performance by reducing the ASR to below 0.5\%  on both ImageNet and MSCOCO datasets in the vast majority of attack scenarios, significantly outperforming most existing defense baselines. Notably, when defending against the sophisticated BadCLIP attack, existing baseline methods exhibit limited efficacy: only PAR demonstrates partial mitigation capabilities yet still retains unacceptably high residual ASR e.g., >15\%. In contrast, InverTune achieves comprehensive defense by suppressing ASR to 0.68\% without compromising model utility. Specifically, (1) For image classification, InverTune reduces ASR from 98.36\% to 0.49\%, representing a 17.29 percentage-point improvement over PAR's 17.78\% residual ASR; (2) For cross-modal retrieval tasks, it decreases ASR from 99.28\% to 0.68\%, outperforming PAR by 15.79 percentage points (16.47\% vs 0.68\%). 

\noindent\textbf{Model Performance}.
We notice that InverTune maintains exceptional preservation of model utility across diverse scenarios compared to baselines. Empirical evaluations across 12 experimental configurations with 2 tasks $\times$ 6 attack methods reveal that our method achieves either the highest (6 cases) or second-highest (5 cases) CA. Though trailing PAR by 0.68\% in ImageNet's Blended scenario (53.50\% vs 54.18\%), this minor gap is statistically insignificant compared to its 17.29\% ASR advantage (BadCLIP).
Beside, when defending against the BadEncoder attack on ImageNet, CleanerCLIP achieves 55.98\% CA slightly higher than  InverTune's 55.84\%. However, the high ASR of CleanerCLIP with 19.71\% demonstrates the weak defense capability.
Moreover, we highlight that InverTune achieves a superior security-performance trade-off which is raised by the correct backdoor information identification.

\begin{figure}[t]
  \centering
  \begin{subfigure}[b]{0.49\linewidth}
    \centering
    \includegraphics[width=\linewidth]{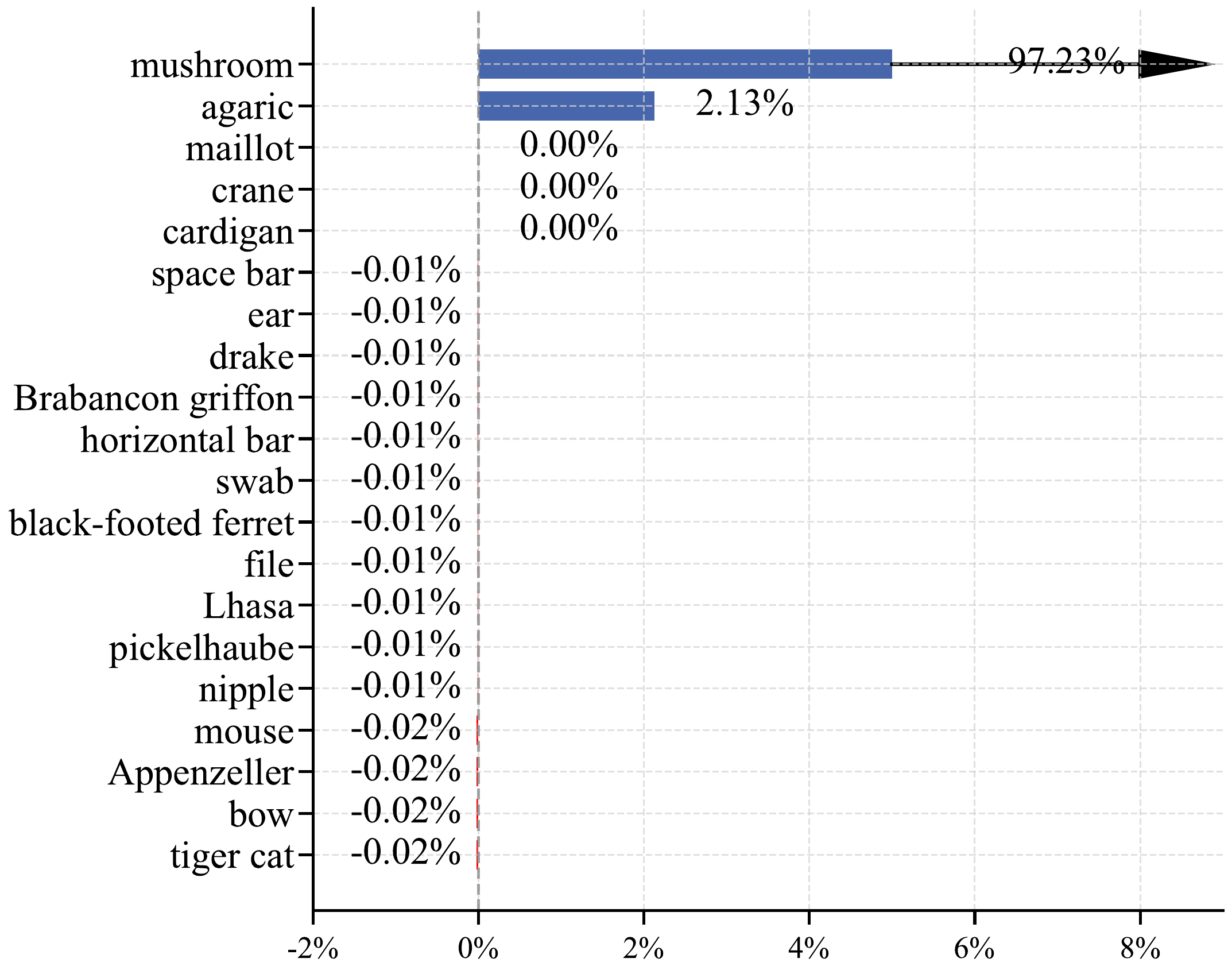}
    \caption{Top 20 classes with increased prediction frequency.}
    \label{subfig:target_label}
  \end{subfigure}
  \hfill
  \begin{subfigure}[b]{0.48\linewidth}
    \centering
    \includegraphics[width=\linewidth]{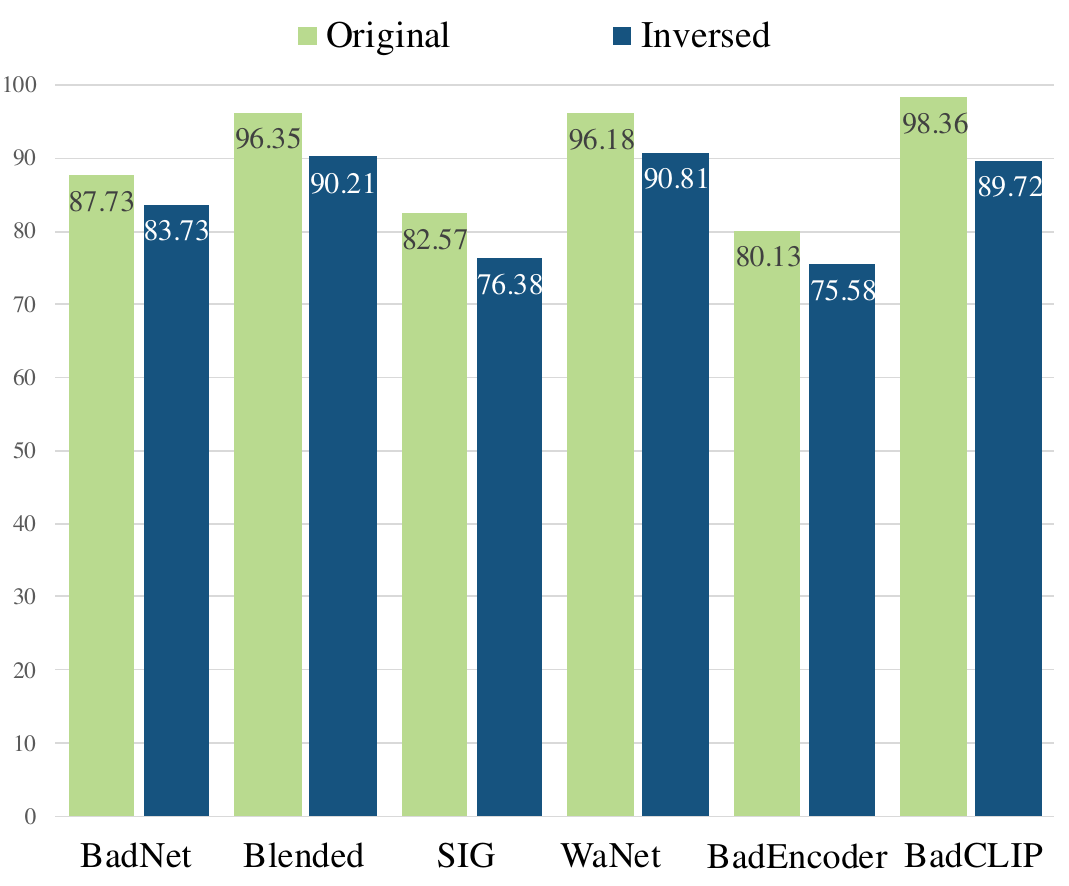}
    \caption{ASR of inverted and original trigger.}
    \label{subfig:trigger_ASR}
  \end{subfigure}
  \caption{Results of backdoor target identification and trigger inversion.}
  \label{fig:two_subfigures}
\end{figure}

\noindent\textbf{Backdoor label Identification and Inversion}.
InverTune consists of two important steps: target identification and trigger inversion.  To demonstrate this effectiveness, we exhibit the corresponding results. 
For step 1, we apply universal adversarial perturbations to clean examples and feed them the compromised model with ``mushroom'' as the designated target class. As demonstrated in Figure~\ref{subfig:target_label}, we observe dramatic distribution shifts in prediction frequencies. Specifically, only two categories exhibit notable increases:  ``mushroom'' shows a 97.23\% surge in classification frequency compared to clean samples, while ``agaric'', (a mushroom subspecies sharing similar visual characteristics), experiences a marginal 2.13\% rise. 
This divergence distribution reveals that adversarial perturbations are effectively utilized to identify the target label.  For the second step, we reconstruct trigger patterns. Here, we argue that the trigger we construct is to activate backdoor pathways for defense without requiring physical trigger replication. As shown in  Figure \ref{subfig:trigger_ASR}, inverted triggers achieve similar attack behavior alignment with original patterns, meaning the inverted trigger largely mimics the attack behavior of the real trigger. 

\begin{table*}[t]
  \caption{Influence of $\lambda$ Parameters on Reverse-Engineered Trigger ASR.}
  \label{hyperparameters}
  \centering
  \begin{tabular}{@{}c|cccc|cccc|cccc|cccc@{}}
    \toprule
    \multirow{2}{*}{\textbf{Attacks}} & \multicolumn{4}{c|}{$\lambda_1$} & \multicolumn{4}{c|}{$\lambda_2$} & \multicolumn{4}{c|}{$\lambda_3$} & \multicolumn{4}{c}{$\lambda_4$} \\
    \cmidrule(lr){2-5} \cmidrule(lr){6-9} \cmidrule(lr){10-13} \cmidrule(lr){14-17}
    & 1.0 & 5.0 & 10.0 & 20.0 & 0.1 & 0.5 & 1.0 & 5.0 & 0.5 & 1.0 & 5.0 & 10.0 & 0.005 & 0.01 & 0.05 & 0.1 \\
    \midrule
    BadNet & 51.97 & 83.73 & 84.84 & 87.67 &
    87.11 & 83.73 & 83.86 &60.86 & 
    90.52 & 83.73 & 84.53 & 65.42 &
    81.70 & 83.73 & 48.75 & 43.32 \\
    Blended & 37.09 & 90.21 & 92.09 & 92.62 &
    86.72 & 90.21 & 89.57 & 62.33 &
    91.83 & 90.21 & 53.17 & 33.38 &
    93.97 & 90.21 & 20.02 & 10.02\\
    SIG & 39.84 & 76.38 & 78.84 & 79.29 & 
    73.62 & 76.38 & 71.18 & 71.74 &
    80.02 & 76.38 & 41.74 & 20.08 &
    80.10 & 76.38 & 12.44 & 0.05\\
    WaNet & 60.81 & 90.81 & 93.96& 89.83 & 
    71.53 & 90.81 & 87.15 & 81.04 &
    92.16 & 90.81 & 78.37 & 37.02 &
    92.38 & 90.81 & 30.06 & 20.03 \\
    BadEncoder & 68.72 & 75.58 & 77.43 & 78.59 &
    75.13 & 75.58 & 70.76 & 70.23 &
    75.04 & 75.58 & 66.64 & 65.38 &
    77.13 & 75.58 & 65.23 & 62.52 \\
    BadCLIP & 73.38 & 89.72 & 87.38 & 89.48 & 
    79.89 & 89.72 & 65.08 & 57.17 &
    69.10 & 89.72 & 73.54 &69.74 &
    91.13 & 89.72 & 59.86 & 20.83 \\
    \bottomrule
  \end{tabular}
\end{table*}

\begin{table*}[t]
  \caption{Comparison of universal adversarial perturbation (UAP) and inverted trigger (InvT) for the Activation Tuning.} 
  \label{tab:ablation_1}
    \setlength{\tabcolsep}{4pt} 
  \renewcommand{\arraystretch}{0.9}
  \centering
  \begin{tabular}{@{}lccccccccccccc@{}}
    \toprule
    \multicolumn{2}{c}{\multirow{2}{*}{\textbf{Methods}}} & \multicolumn{2}{c}{\textbf{BadNet}} & \multicolumn{2}{c}{\textbf{Blended}}  & \multicolumn{2}{c}{\textbf{SIG}} & \multicolumn{2}{c}{\textbf{WaNet}} & \multicolumn{2}{c}{\textbf{BadEncoder}}  & \multicolumn{2}{c}{\textbf{BadCLIP}} \\
    \cmidrule(lr){3-4} \cmidrule(lr){5-6} \cmidrule(lr){7-8} \cmidrule(lr){9-10} \cmidrule(lr){11-12} \cmidrule(lr){13-14}  
    & & CA $\uparrow$ & ASR $\downarrow$ & CA $\uparrow$ & ASR $\downarrow$ & CA $\uparrow$ & ASR $\downarrow$ & CA $\uparrow$ & ASR $\downarrow$ & CA $\uparrow$ & ASR $\downarrow$ & CA $\uparrow$ & ASR $\downarrow$ \\
    \midrule
    \multirow{2}{*}{Top-1} & UAP & 55.55 & 23.24 & 53.46 & 89.92 & \textbf{54.27 }& 58.91 & 52.25 & 25.74 & 52.81 & 67.13 & 52.02 & 54.33\\
    & InvT &\textbf{56.12} & \textbf{0.02} & \textbf{53.50} & \textbf{0.14} & \textbf{54.27} & \textbf{0.03} & \textbf{54.76} & \textbf{0.09} & \textbf{55.84} & \textbf{0.02} & \textbf{55.25} & \textbf{0.49}\\
    \midrule
    \multirow{2}{*}{Top-3} & UAP & 76.76 & 47.39 & 74.99 & 95.49 & 75.71 & 76.91 & 73.64 & 50.86 & 74.67 & 69.76 & 73.42 & 71.15\\
    & InvT & \textbf{77.24} & \textbf{0.20} & \textbf{75.35} & \textbf{0.74} & \textbf{75.92} & \textbf{0.10} & \textbf{75.92} & \textbf{0.40} & \textbf{77.05} & \textbf{0.20} & \textbf{76.45} & \textbf{1.17}\\
    \midrule
    \multirow{2}{*}{Top-5} & UAP & 83.63 & 58.98 & 82.02 & 96.75 & 82.84 & 82.16 & 81.05 & 60.76 & 81.73 & 71.04 & 80.80 & 76.18\\
    & InvT &\textbf{84.10} & \textbf{0.46} & \textbf{82.54} & \textbf{1.47} & \textbf{83.04} & \textbf{0.20} & \textbf{83.04} & \textbf{0.77} & \textbf{83.87} & \textbf{0.44} & \textbf{83.35} & \textbf{1.67}\\
    \midrule
    \multirow{2}{*}{Top-10} & UAP & 90.11 & 72.94 & 89.01 & 98.01 & 89.53 & 87.85 & 88.33 & 72.45 & 88.76 & 72.85 & 88.13 & 81.88\\
    & InvT &\textbf{90.56} & \textbf{0.95} & \textbf{89.52} & \textbf{3.42} & \textbf{89.80} & \textbf{0.41} & \textbf{89.80} & \textbf{1.71} & \textbf{90.37} & \textbf{0.91} & \textbf{90.00} & \textbf{2.60}\\ \bottomrule
  \end{tabular}
\end{table*}

\begin{figure}[t]
  \centering
  \includegraphics[width=0.9\linewidth, height=0.5\linewidth]{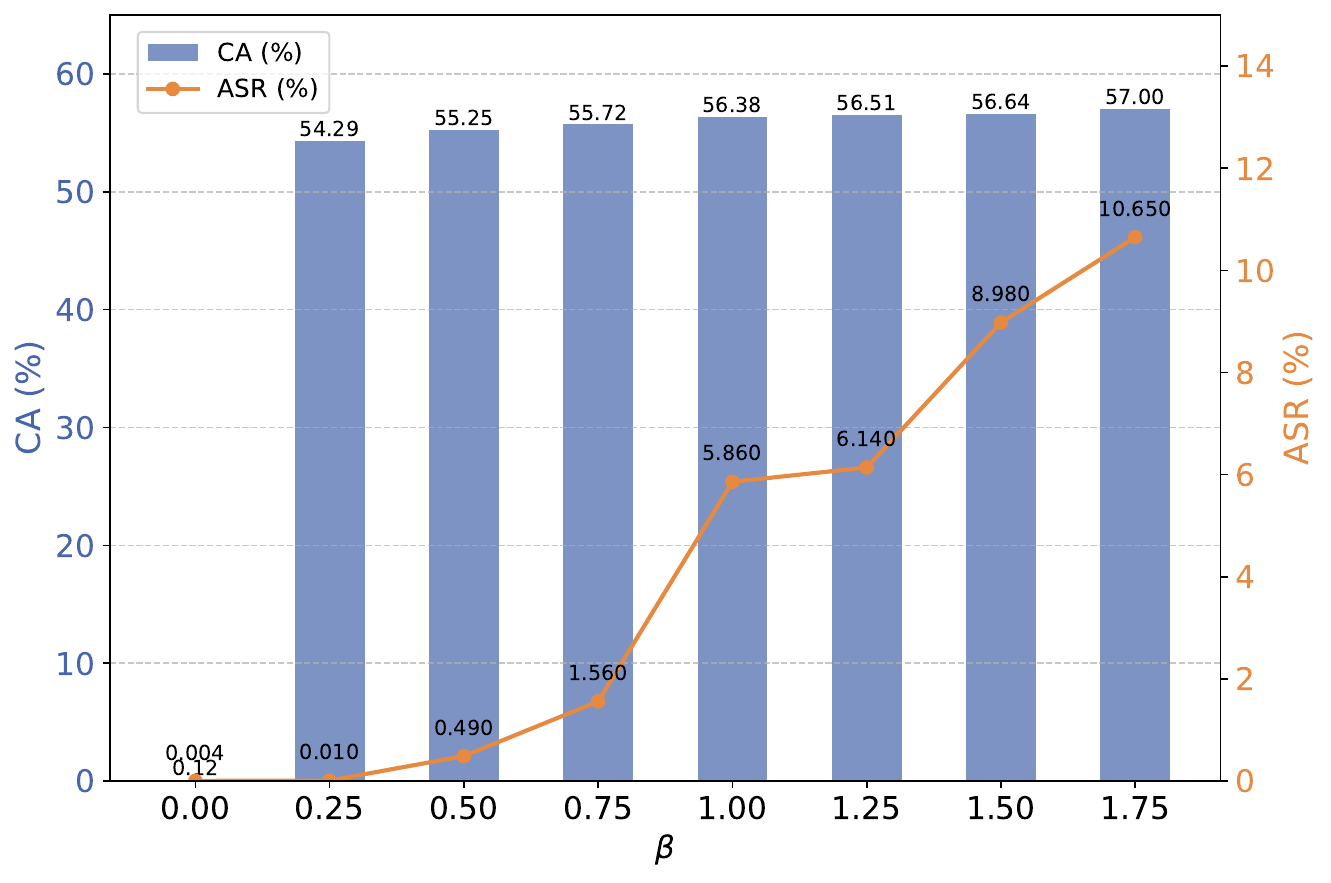}
  \vspace{-1em}
  \caption{Influence of $\beta$ on InverTune's Defense Effectiveness under BadClip Attack Scenario.}
  \label{fig:step3}
\end{figure}

\subsection{Influence of hyperparameters}
In this section, we study the influence of different hyperparameters. As formulated in Equation~\eqref{eq:inver_loss}, the coefficients $\lambda_1$-$\lambda_4$ control the relative importance of four loss components during backdoor inversion, while $\beta$ in Equation~\eqref{eq:finetune_loss} governs the trade-off between model cleanliness and usability during the elimination phase.

Our experiments show several important patterns in hyperparameter sensitivity.
For the inversion-related hyperparameters (see in Table~\ref{hyperparameters}), we observe that $\lambda_1$, which weights the contrastive learning loss, produces significantly improved ASR when increased, though with diminishing returns beyond $\lambda_1 > 5.0$ due to deteriorating visual quality of the inverted triggers. The visual feature consistency term controlled by $\lambda_2$ demonstrates a clear sweet spot, where insufficient weighting ($\lambda_2 = 0.1$) fails to achieve effective attacks, particularly on SIG, WaNet, and BadCLIP, while excessive emphasis ($\lambda_2 > 1.0$) degrades ASR by over-constraining the feature space. Optimal visual quality and attack effectiveness are achieved with $\lambda_3 = 1.0$ and $\lambda_4 = 0.01$, which properly balance trigger stealthiness and functionality. Excessive values of $\lambda_3$ and $\lambda_4$ shift the focus of the inversion process towards trigger size optimization, thereby compromising the adversarial effectiveness of the inverted triggers.

The elimination phase analysis (see in Figure~\ref{fig:step3}) shows the critical role of $\beta$ in balancing security and utility. The extreme case of $\beta=0$, which completely prioritizes backdoor removal, reduces both CA (0.12\%) and ASR (0.004\%) to near-zero levels, validating the necessity of the usability term in Equation~\ref{e3}. As $\beta$ increases, we observe distinct patterns: CA shows stable improvement that plateaus when $\beta > 0.50$, while ASR exhibits more dramatic growth, particularly in the range $\beta \in [0.75,1.0]$ where it increases from 1.560\% to 5.860\%. Our selected value $\beta=0.5$ achieves an effective balance, maintaining ASR at 0.49\% while preserving 55.25\% CA, demonstrating both the stability of InverTune and the effectiveness of our loss formulation.

\subsection{Ablation Study}
Our analysis in Section~\ref{sec:step1} reveals behavioral distinctions and connections between adversarial and backdoor-triggered samples in compromised models. While both input types induce target-class misclassification, they exploit fundamentally different model vulnerabilities. This mechanistic necessitates our novel trigger inversion approach to specifically isolate and neutralize backdoor artifacts rather than relying solely on adversarial patterns.

To empirically validate this requirement, we conduct an ablation study comparing the complete InverTune framework (InvT) against a variant (UAP) that directly fine-tunes using first-stage adversarial perturbations while omitting trigger inversion. As shown in Table \ref{tab:ablation_1}, InvT demonstrates overwhelming superiority across all metrics. When k=1, 3, 5, and 10, the average ASR of InvT is 0.13\%, 0.46\%, 0.84\%, and 1.67\% respectively, significantly outperforming UAP with 53.21\%, 68.59\%, 74.31\%, 81.00\%.  The performance gap stems from UAP's fundamental limitation: while adversarial fine-tuning enhances noise robustness and marginally reduces surface-level ASR, it fails to address deeper backdoor information. Moreover, InvT simultaneously preserves superior CA through targeted backdoor pathway disruption compared to indiscriminate adversarial examples. These experiments demonstrate the effectiveness and necessity of the inversion step in InverTune, as it not only enhances backdoor removal but also better preserves the model’s usability.

\subsection{Backdoor Configuration}\label{sec:targetlabel}
Section \ref{sec:performance} presents a comprehensive evaluation of InverTune's effectiveness. In this sections, we conduct in-depth analyses of the defense mechanism across different dimensions, including target labels and model architectures.  To save resources, we mainly focus on \textbf{BadCLIP}, which represents the most advanced attack and poses the most significant challenge to defenses.

\noindent\textbf{The impact of target label.} To assess the generalizability of InverTune across diverse attack targets, we further set ``banana'', ``lemon'', and ``ski'' as the target label and train BadCLIP attack models with distinct trigger patterns.  As illustrated in Table \ref{lableA}, baseline like FT and CleanCLIP remain vulnerable to BadCLIP attacks regardless of target label variations. More advanced defenses such as CleanerCLIP and PAR exhibit notable performance fluctuations: CleanerCLIP's effectiveness decreases from 16.16\% to 25.36\% and PAR's from 11.72\% to 36.07\% when switching from ``ski'' to ``lemon''. In contrast, InverTune maintains consistent defensive capabilities, achieving superior performance in both ASR ($\approx$ 1\%) and CA metrics across all target labels. These results demonstrate InverTune's robust ability to identify and neutralize backdoor threats regardless of the target label selection.

\begin{table}[t]
  \caption{Performance comparison of InverTune and baseline defenses against BadCLIP under different target labels.}
  \label{lableA}
  \centering
  \setlength{\tabcolsep}{4pt} 
  \renewcommand{\arraystretch}{1.0} 
  \resizebox{\columnwidth}{!}{%
    \begin{tabular}{@{}lc|cccccc@{}}
      \toprule
      \multirow{2}{*}{} & \multirow{2}{*}{\textbf{Target Label}} & \multicolumn{2}{c}{\textbf{Banana}} & \multicolumn{2}{c}{\textbf{Lemon}} & \multicolumn{2}{c}{\textbf{Ski}} \\
      \cmidrule(lr){3-4} \cmidrule(lr){5-6} \cmidrule(lr){7-8}  
      &  & CA $\uparrow$ & ASR $\downarrow$ & CA $\uparrow$ & ASR $\downarrow$ & CA $\uparrow$ & ASR $\downarrow$ \\
      \midrule
      & No Defense & 58.20 & 98.16 & 58.11 & 97.16 & 58.31 & 98.46 \\
      & FT         & \underline{54.77} & 83.14 & \underline{54.93} & 89.65 & \underline{54.34} & 79.70 \\
      & CleanCLIP  & 53.48 & 74.85 & 54.50 & 72.82 & 53.94 & 77.75 \\
      & CleanerCLIP & 52.09 & 20.41 & 51.69 & \underline{25.36} & 51.67 & 16.16 \\
      & PAR        & 53.64 & \underline{17.65} & 53.91 & 36.07 & 53.62 & \underline{11.72} \\
      & InverTune (Ours) & \textbf{57.01} & \textbf{1.14} &\textbf{ 55.81} & \textbf{1.01} &\textbf{ 56.93} & \textbf{1.51} \\   
      \bottomrule 
    \end{tabular}%
  }
\end{table}

\noindent\textbf{The impact of model structure.}
To assess the generalizability of InverTune on architectures, we evaluate its performance across diverse model architectures including RN101, ViT-B/16, and ViT-B/32. As shown in Table \ref{table_model_structure}, architectural transitions significantly impact defense performance. Notably, baselines exhibit substantial performance fluctuations across different architectures. For instance, PAR shows severe performance degradation when transitioning from RN101 to ViT-B/32, with ASR increasing dramatically from 1.17\% to 76.37\% and CA declining from 55.60\% to 50.82\%. Similarly, CleanerCLIP's effectiveness varies considerably, with ASR ranging from 3.25\% to 64.60\% across different architectures. In contrast, InverTune exhibits remarkable stability and superior defensive capability across all evaluated architectures, maintaining an average ASR of merely 1.22\% under BadCLIP attacks while preserving competitive CA. This consistent performance across both CNN-based (RN50, RN101) and Transformer-based (ViT-B/32, ViT-B/16) architectures validates its architectural robustness and generalizability. This architecture-agnostic effectiveness originates from InverTune's backdoor inversion paradigm, which directly targets fundamental cross-modal activation patterns rather than architecture-specific features.

\begin{table}[t]
  \caption{Performance comparison of defense methods across different model architectures.}
  \vspace{-0.1em}
  \centering
  \label{table_model_structure}
  \setlength{\tabcolsep}{4pt} 
  \renewcommand{\arraystretch}{1.0} 
  \resizebox{\columnwidth}{!}{%
    \begin{tabular}{@{}lc|cccccc@{}}
      \toprule 
      \multirow{2}{*}{} & \multirow{2}{*}{\textbf{Backbone}} & \multicolumn{2}{c}{\textbf{RN101}} & \multicolumn{2}{c}{\textbf{ViT-B/16}} & \multicolumn{2}{c}{\textbf{ViT-B/32}} \\
      \cmidrule(lr){3-4} \cmidrule(lr){5-6} \cmidrule(lr){7-8}  
      &  & CA $\uparrow$ & ASR $\downarrow$ & CA $\uparrow$ & ASR $\downarrow$ & CA $\uparrow$ & ASR $\downarrow$ \\
      \midrule
      & No Defense & 59.17 & 83.17 & 66.78 & 99.90 & 60.97 & 99.23 \\
      & FT         & \textbf{56.85} & 58.29 & \textbf{63.01} & 83.75 & \underline{54.72} & 91.61 \\
      & CleanCLIP  & \underline{56.14} & 42.53 & \underline{61.91} & 80.33 & 53.16 & 79.36 \\
      & CleanerCLIP & 52.76 & 3.25 & 58.81 & 31.17 & 53.64 & \underline{64.60} \\
      & PAR        & 55.60 & \underline{1.17} & 57.98 & \underline{18.14} & 50.82 & 76.37 \\
      & InverTune (Ours) & 55.76 & \textbf{1.00} & 59.80 & \textbf{0.09 }& \textbf{54.83} & \textbf{0.17} \\   
      \bottomrule 
    \end{tabular}%
  }
\end{table}

\section{Related Work}

\subsection{Backdoor Attacks in MCL}

Traditional backdoor attacks, such as BadNet~\cite{gu2017badnets}, Blended~\cite{chen2017targeted}, SIG~\cite{barni2019new} and TrojanNet~\cite{tang2020embarrassingly}, originally target unimodal neural networks but can be adapted to compromise Multimodal Contrastive Learning models through data poisoning. However, recent MCL-specific attacks exploit cross-modal interactions more effectively. Carlini et al.~\cite{carlini2021poisoning} show that minimal data poisoning can introduce severe vulnerabilities. BadEncoder~\cite{jia2022badencoder} targets self-supervised learning by poisoning pre-trained image encoders, causing downstream classifiers to inherit backdoor behaviors while maintaining model accuracy. GhostEncoder~\cite{wang2024ghostencoder} introduces a dynamic invisible backdoor using image steganography to embed hidden triggers into benign images. 
Notably, BadCLIP~\cite{liang2024badclip} introduces a dual-embedding framework that aligns poisoned samples with target features, creating natural-looking triggers resistant to standard defenses. Adding to this threat landscape, Bai et al.~\cite{bai2024badclip} propose a prompt-based backdoor attack that manipulates both image and text encoders using learnable triggers and trigger-aware prompts. These approaches highlight the diverse strategies employed in backdoor attacks and the urgent need for effective defenses.

\subsection{Backdoor Defenses in MCL}

Backdoor defenses in MCL involve both detection and mitigation strategies. DECREE~\cite{feng2023detecting} focuses on identifying backdoors but lacks effective mechanisms for removal. While SSL-Cleanse~\cite{zheng2024ssl} is designed for self-supervised learning, it not only detects backdoors but also incorporates a purification process to mitigate them. Fine-tuning-based approaches, such as CleanCLIP~\cite{bansal2023cleanclip}, PAR~\cite{singh2024perturb}, and CleanerCLIP~\cite{DBLP:journals/corr/abs-2409-17601}, attempt to remove backdoors by re-learning representations or leveraging counterfactual augmentations. However, these methods may require large clean datasets or introduce performance trade-offs.
ABD~\cite{kuang2024adversarial} creatively leverages adversarial examples to approximate backdoor samples but faces challenges in maintaining clean accuracy. Pre-training defenses, such as RoCLIP~\cite{yang2023robust} and SafeCLIP~\cite{yang2023better}, mitigate backdoors by filtering poisoned data during pre-training. However, their effectiveness relies on access to the pre-training process, making them unsuitable for scenarios where only a trained model is available. Our method aims to effectively eliminate backdoor threats in multimodal contrastive learning models while preserving their original performance and generalization capabilities.

\section{Conclusion}
In this paper, we present InverTune, a novel backdoor defense framework for large-scale multimodal contrastive learning models. Our approach integrates three key components: adversarial-based target label identification, gradient-guided trigger inversion, and activation-aware fine-tuning. Extensive evaluations on multiple datasets demonstrate that InverTune achieves state-of-the-art defensive performance across diverse attack scenarios, consistently reducing attack success rates while maintaining model utility. Our framework significantly enhances the robustness of multimodal models against backdoor threats, providing a practical solution for real-world applications.

\clearpage

\bibliographystyle{ACM-Reference-Format}
\bibliography{main}

\appendix
\section{Introduction and Configurations of Different Backdoor Attacks}
\label{sec2}
\subsection{Backdoor Attacks Settings}

For all six types of attacks, we adopt a 500K subset of the CC3M dataset~\cite{sharma2018conceptual} as the fine-tuning dataset. All attacks target the class “mushroom.” For attacks that require textual descriptions, we construct them by collecting 131 mushroom-related captions from the CC3M dataset and randomly assigning them to the poisoned image samples as their corresponding text descriptions.

\begin{itemize}
\item In the BadNet~\cite{gu2017badnets} attack, we adopt a $16 \times 16$ patch filled with Gaussian noise sampled from a standard normal distribution as the trigger, which is fixed to the bottom-right corner of the clean images.

\item In the Blended~\cite{chen2017targeted} attack, we generate a trigger image of the same size as the input image using a uniform distribution. We set the transparency of the trigger image to 0.2 and blend it with the clean image, whose transparency is set to 0.8.

\item In the SIG~\cite{barni2019new} attack, sinusoidal noise is generated along the horizontal axis of the image, creating vertical stripes. For each pixel along the width, noise is injected using a sinusoidal function with a frequency of 6 cycles per image width. The noise amplitude is scaled to $60/255$ to stay within a suitable range. This perturbation is applied uniformly to all RGB channels. After adding the noise, pixel values are clipped to $[0, 1]$ to ensure validity. 

\item In the WaNet~\cite{nguyen2021wanet} attack, we apply a warping transformation to the image using a distortion grid. Following the original implementation, we generate the grid by interpolating a noise tensor to match the image resolution. The grid is then scaled and clipped to $[-1, 1]$ for compatibility with grid sampling. The warping is performed using bilinear interpolation, introducing subtle but adversarial distortions. 

\item For the BadEncoder~\cite{jia2022badencoder} attack, we follow the original methodology, where the visual encoder is fine-tuned to embed backdoor triggers while preserving its functionality on clean samples. Unlike the original implementation, we replace the trigger from the official repository with a $16 \times 16$ pure white image to ensure a fair comparison with other attacks. This attack is distinct in that it does not require constructing textual descriptions or setting a poisoning rate. Instead, it directly fine-tunes the visual encoder using a reference dataset and a shadow dataset. 

\item For the BadCLIP~\cite{liang2024badclip} attack, following their provided code, we first optimize the patch based on the ``mushroom'' label. After obtaining the patch, we perform Dual-Embedding injection attack on the clean CLIP model.
\end{itemize}

For all the attacks described above, we start from the CLIP model pretrained by OpenAI~\cite{radford2021learning}, and fine-tune it to obtain a poisoned CLIP model with learning rate 1e-6, batch size 128, and 10 training epochs.

\subsection{Visualization of Trigger Patterns}

\begin{figure*}[t!] 
  \centering
  \includegraphics[width=.95\linewidth]{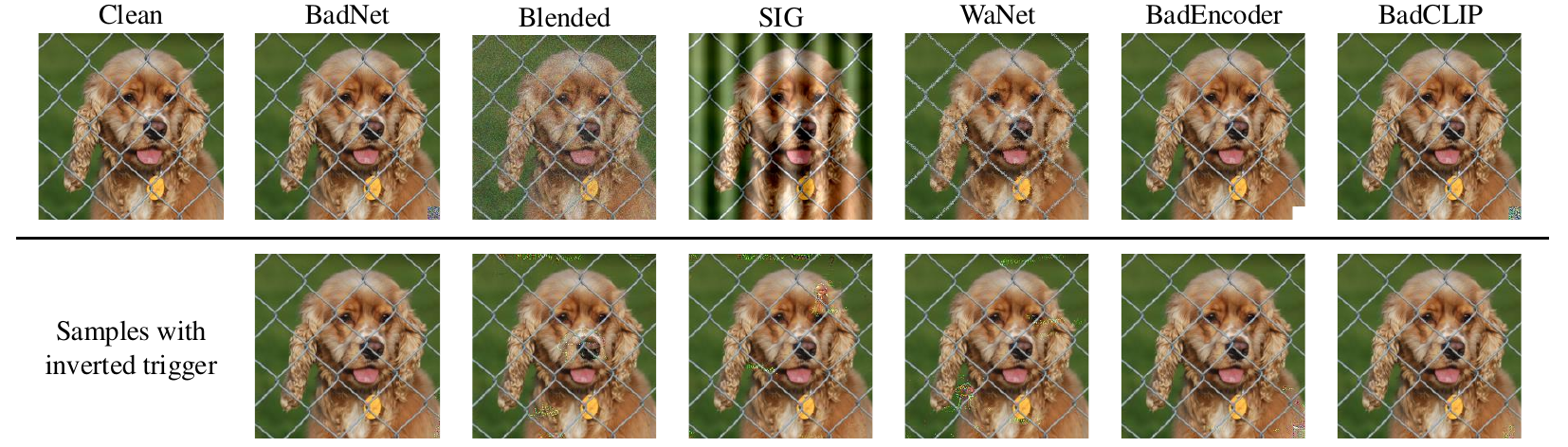}
  \caption{Backdoor sample examples and visualization of trigger inversion effects.}
  \label{fig:comparison}
\end{figure*}

Regarding the attacks mentioned in this paper, in addition to the introduction above, we also present them in the first row of Figure~\ref{fig:comparison}. 

\section{Baseline Defense Settings}
\label{sec3}
In this section, we provide a detailed description of the experimental settings for the four baseline methods discussed in the Main Text.

All the defense methods use subsets of the CC3M dataset~\cite{sharma2018conceptual} in their original setups, though the exact number of samples varies slightly. For fair comparison, we standardize the training data by using a fixed subset of 500K samples across all methods.

\begin{itemize}

\item The fine-tuning method (FT), first introduced by CleanCLIP~\cite{bansal2023cleanclip}, involves fine-tuning the model with a multimodal contrastive loss on a clean dataset. In our experiments, we use the official implementation provided by CleanCLIP, with a learning rate of 4.5e-6, warmup steps of 50, batch size of 64, and 10 training epochs.

\item CleanCLIP~\cite{bansal2023cleanclip} extends FT by adding a self-supervised loss term. Following its original setup, we set the weights of the self-supervised loss term and the contrastive loss term to 1, with other hyperparameters remaining the same as those in FT.

\item PAR~\cite{singh2024perturb} adopts a custom learning rate schedule. However, due to the increased size of the fine-tuning dataset, the original setting does not reproduce the reported performance. Therefore, in our experiments, we modify the start learning rate to 3e-6 and the peak learning rate to 5e-6, while keeping all other parameters consistent with the original setup.

\item CleanerCLIP~\cite{DBLP:journals/corr/abs-2409-17601} is implemented based on CleanCLIP~\cite{bansal2023cleanclip}. We follow its original setup, using a batch size of 64 and training for 10 epochs with the AdamW optimizer. The learning rate is linearly warmed up over 10,000 steps, and a weight decay of 0.1 is applied. The Adam momentum factor and RMSProp factor are set to 0.9 and 0.999, respectively, with an epsilon of 1e-8. The base learning rate is set to 4.5e-6.

\end{itemize}

\section{Implementation Details of InverTune}
\label{sec4}


\subsection{Backdoor Label Identification}
\label{sec:label}

The first step of InverTune is to identify the target category for the backdoor attack. To achieve this, we leverage the 1,000 classes from ImageNet-1K~\cite{ILSVRC15} and combine them with predefined templates to construct text prompts.

These categories are derived from WordNet~\cite{miller1995wordnet}, a lexical database that structures words into a hierarchical network based on their semantic relationships. The ImageNet-1K classes encompass a remarkably diverse array of objects, spanning nearly all aspects of the physical world. These include animals (e.g., tiger, goldfish, hummingbird), everyday objects (e.g., laptop, toaster, umbrella), vehicles (e.g., fire truck, sports car, airplane), architectural structures (e.g., lighthouse, suspension bridge, pagoda), and various tools and instruments (e.g., screwdriver, stethoscope, cello).

Given their extensive coverage, these 1,000 categories serve as well-suited candidates for identifying the target labels in backdoor attacks. Their diversity ensures a broad spectrum of potential backdoor targets, making them highly relevant for identifying and mitigating threats in multimodal contrastive learning models. Additionally, the hierarchical nature of WordNet provides a strong semantic foundation, facilitating precise and meaningful target label selection.

Furthermore, Even if the attacker chooses a target category outside ImageNet-1K, a semantically similar class likely exists within it due to WordNet’s hierarchy. This ensures our defense remains effective, as the attacker’s target can still be meaningfully mapped to an existing label, maintaining robustness against unexpected attacks.

\subsection{Trigger Inversion}

Algorithm~\ref{alg:trigger_inversion} generates high-fidelity backdoor trigger reconstructions while maintaining visual subtlety. The four loss components work together to achieve this, as detailed in Section 3.2 of the Main Text.

The trigger reconstruction process typically converges within a few hundred iterations, significantly faster than training a backdoor from scratch. This efficiency stems from directly optimizing in CLIP's embedding space rather than attempting to model the backdoor through proxy tasks or surrogate networks.

Importantly, our approach supports a wide range of backdoor implementations beyond the standard patch-based triggers. The mask-pattern formulation can reconstruct complex, spatially distributed triggers and even global transformations. The clamp operation on the trigger pattern (line 26, 27) ensures the reconstructed values remain within CLIP's preprocessing bounds, producing realistic images that can be directly used in subsequent defense strategies.

The reconstructed trigger serves as a critical component for our overall defense framework, enabling us to analyze backdoor behavior and develop targeted mitigation strategies in the activation tuning stage. By reproducing the backdoor's trigger, we can effectively probe the model's internal representations to identify compromised components.

\begin{figure*}[t!]
\centering
\begin{minipage}{\textwidth}
\begin{algorithm}[H]
\caption{Dual-Space Trigger Inversion for Multimodal CLIP Backdoors}
\label{alg:trigger_inversion}
\begin{algorithmic}[1]
\State \textbf{Input:} Suspected backdoored CLIP model $F$ with image encoder $E_I$ and text encoder $E_T$; Clean images $\mathcal{X} = \{x_1, x_2, \ldots, x_n\}$; Target text label $y_t$ identified from Step 1; Number of steps $T$; Loss weights $\lambda_1, \lambda_2, \lambda_3, \lambda_4$
\State \textbf{Output:} Inverted trigger mask $m$ and pattern $t_{\text{img}}$

\State Initialize mask $m \gets$ random tensor in range $[0,1]$ of shape $3 \times 224 \times 224$
\State Initialize trigger pattern $t_{\text{img}} \gets$ random tensor of shape $3 \times 224 \times 224$
\State $\theta \gets \{m, t_{\text{img}}\}$ \Comment{Parameters to optimize}
\State Initialize optimizer with learning rate $\alpha$

\State Precompute all text embeddings for available classes $\{y_1, y_2, \ldots, y_N\}$:
\State $\quad E_T(y_j) \gets$ normalized text embeddings for each class $j \in \{1,\ldots,N\}$

\For{$\text{step} = 1$ \textbf{to} $T$}
    \State Sample a batch of clean images $\{x_1, x_2, \ldots, x_b\} \subseteq \mathcal{X}$
    
    \State Generate poisoned samples: $\tilde{x}_i = m \odot t_{\text{img}} + (1-m) \odot x_i$ for $i \in \{1,\ldots,b\}$
    
    \State Compute image embeddings: $E_I(\tilde{x}_i) \gets F_I(\tilde{x}_i)$
    \State Normalize embeddings: $E_I(\tilde{x}_i) \gets \frac{E_I(\tilde{x}_i)}{\|E_I(\tilde{x}_i)\|_2}$
    
    \State \Comment{Calculate the four loss components}
    
    \State \Comment{1. Cross-Modal Alignment Loss via InfoNCE}
    \State $\mathcal{L}_{\text{align}} \gets -\frac{1}{b}\sum_{i=1}^{b}\log\frac{\exp(\text{sim}(E_I(\tilde{x}_i), E_T(y_t))/\tau)}{\sum_{j=1}^{N}\exp(\text{sim}(E_I(\tilde{x}_i), E_T(y_j))/\tau)}$
    
    \State \Comment{2. Embedding Space Preservation Loss}
    \State $\mathcal{L}_{\text{emb}} \gets \frac{1}{b}\sum_{i=1}^{b}\|E_I(\tilde{x}_i) - E_I(x_i)\|_2$
    
    \State \Comment{3. Visual Similarity Loss}
    \State $\mathcal{L}_{\text{sim}} \gets \frac{1}{b}\sum_{i=1}^{b}(1 - \text{SSIM}(\tilde{x}_i, x_i))$
    
    \State \Comment{4. Trigger Sparsity Loss}
    \State $\mathcal{L}_{\text{mask}} \gets \|m\|_1$
    
    \State \Comment{Combined Loss}
    \State $\mathcal{L}_{\text{inver}} \gets \lambda_1\mathcal{L}_{\text{align}} + \lambda_2\mathcal{L}_{\text{emb}} + \lambda_3\mathcal{L}_{\text{sim}} + \lambda_4\mathcal{L}_{\text{mask}}$
    
    \State Update parameters: $\theta \gets \theta - \alpha\nabla_{\theta}\mathcal{L}_{\text{inver}}$
    
    \State Clamp mask: $m \gets \text{clamp}(m, 0, 1)$
    \State Clamp trigger: $t_{\text{img}} \gets \text{clamp}(t_{\text{img}}, -1.7922, 2.1461)$ \Comment{CLIP normalization bounds}
\EndFor

\State \Return $m, t_{\text{img}}$
\end{algorithmic}
\end{algorithm}
\end{minipage}
\end{figure*}

\subsection{Activation Tuning}

After trigger inversion, we focus on mitigating its impact on the model without compromising normal functionality. Traditional fine-tuning methods applied to the entire network risk degrading the model's critical cross-modal performance, which is essential for multimodal models like CLIP. In contrast, Algorithm~\ref{alg:activation_tuning} introduces a novel activation tuning approach that specifically targets neurons involved in backdoor behavior.

The algorithm operates in three phases: (1) identifying network layers most affected by the backdoor trigger, (2) pinpointing the specific neurons within these layers responsible for the backdoor behavior, and (3) selectively fine-tuning only the identified neurons using a custom loss function. This targeted approach minimizes disruption to the model's cross-modal alignment, which is central to CLIP’s zero-shot prediction capabilities.

By focusing on critical neurons identified through activation analysis, Algorithm~\ref{alg:activation_tuning} offers significant advantages over traditional backdoor mitigation techniques. This selective intervention is more efficient than whole-network fine-tuning, preserving CLIP's core functionality while effectively addressing backdoor pathways.

\begin{figure*}[t!]
\centering
\begin{minipage}{\textwidth}
\begin{algorithm}[H]
\caption{Activation Tuning for Backdoor Mitigation in MCL Models}
\label{alg:activation_tuning}
\begin{algorithmic}[1]
\State \textbf{Input:} Backdoored CLIP model $F$ with encoders $E_I$ and $E_T$; Inverted trigger $(m, t_{\text{img}})$ from Algorithm \ref{alg:trigger_inversion}; Clean inputs $\mathcal{X}$; Set of candidate layers $\mathcal{L}$; Balance parameter $\beta$
\State \textbf{Output:} Fine-tuned CLIP model with neutralized backdoor

\State \textbf{Phase 1: Identify Critical Layers}
\State Compute clean activations $\{A_{\text{clean}}^{l}\}$ for each layer $l \in \mathcal{L}$ using $\mathcal{X}$
\For{each $x \in \mathcal{X}$}
    \State Generate triggered image $\tilde{x} \gets m \odot t_{\text{img}} + (1-m) \odot x$
\EndFor
\State Compute triggered activations $\{A_{\text{triggered}}^{l}\}$ for each layer $l \in \mathcal{L}$

\For{each layer $l \in \mathcal{L}$}
    \State Compute mean clean activation $\mu_{\text{clean}}^{l} \gets \text{mean}(A_{\text{clean}}^{l})$
    \State Compute mean triggered activation $\mu_{\text{triggered}}^{l} \gets \text{mean}(A_{\text{triggered}}^{l})$
    \State Calculate normalized activation difference:
    \State $\text{diff}^{l} \gets \frac{\|\mu_{\text{clean}}^{l} - \mu_{\text{triggered}}^{l}\|_2}{\|\mu_{\text{clean}}^{l}\|_2}$
\EndFor

\State Calculate threshold $\tau \gets \text{mean}(\{\text{diff}^{l}\}) + \text{std}(\{\text{diff}^{l}\})$
\State Identify critical layers $\mathcal{L}_{\text{critical}} \gets \{l \in \mathcal{L} \mid \text{diff}^{l} > \tau\}$

\State \textbf{Phase 2: Identify Critical Neurons}
\For{each layer $l \in \mathcal{L}_{\text{critical}}$}
    \State Compute activation difference $\Delta^{l} \gets |\mu_{\text{clean}}^{l} - \mu_{\text{triggered}}^{l}|$
    \State Apply K-means clustering to $\Delta^{l}$ with $k=2$ clusters
    \State Identify critical cluster $C_{\text{critical}}^{l}$ with largest centroid value
    \State Create neuron mask $M^{l}$ where neurons in $C_{\text{critical}}^{l}$ are set to 1
\EndFor

\State \textbf{Phase 3: Selective Fine-tuning}
\State Create parameter masks based on critical neuron masks $\{M^{l}\}$
\State Initialize fine-tuned model $F' \gets F$ \Comment{Copy of original model}
\State Create optimizer for model parameters with neuron-masked gradients

\For{each training step}
    \State Sample batch of clean images $\{x_1, x_2, ..., x_b\} \subseteq \mathcal{X}$
    \State Generate triggered images $\{\tilde{x}_i = m \odot t_{\text{img}} + (1-m) \odot x_i\}$
    
    \State \Comment{Compute activation alignment loss}
    \State $\mathcal{L}_{\text{activation}} \gets 0$
    \For{each layer $l \in \mathcal{L}_{\text{critical}}$}
        \State Extract activations for clean and triggered inputs: $a_{\text{clean}}^{l}$, $a_{\text{triggered}}^{l}$
        \State Apply neuron mask: $a_{\text{clean}}^{l} \gets a_{\text{clean}}^{l} \odot M^{l}$
        \State Apply neuron mask: $a_{\text{triggered}}^{l} \gets a_{\text{triggered}}^{l} \odot M^{l}$
        \State $\mathcal{L}_{\text{activation}} \gets \mathcal{L}_{\text{activation}} + \|a_{\text{clean}}^{l} - a_{\text{triggered}}^{l}\|_2^2$
    \EndFor
    
    \State \Comment{Compute preservation loss}
    \State With original model $F$, compute $E_I^{\text{orig}}(x_i)$ for each $x_i$
    \State With fine-tuned model $F'$, compute $E_I(x_i)$ for each $x_i$
    \State $\mathcal{L}_{\text{preserve}} \gets \|\text{sim}(E_I(x_i), E_T(y_i)) - \text{sim}(E_I^{\text{orig}}(x_i), E_T(y_i))\|_2^2$
    
    \State \Comment{Combined loss}
    \State $\mathcal{L}_{\text{tune}} \gets \mathcal{L}_{\text{activation}} + \beta \cdot \mathcal{L}_{\text{preserve}}$
    
    \State Compute gradients and apply masked updates to parameters
    \State Update only parameters corresponding to critical neurons
\EndFor

\State \Return Fine-tuned model $F'$
\end{algorithmic}
\end{algorithm}
\end{minipage}
\end{figure*}

\section{Detailed Results of Intermediate Steps in InverTune}
\label{sec5}
\subsection{Target Category Identification Results for Six Attacks}

We present the target class identification results across six distinct attack scenarios, where ``mushroom'' serves as the ground truth target label in all cases. Our analysis reveals systematic and statistically significant increases in the prediction frequency of the target class after adversarial perturbation, with attack-specific variations in magnitude.

\begin{table}[h]
    \centering
    \caption{Top 20 classes with the largest absolute increase under adversarial attack on BadNet-poisoned model.}
    \label{badnet_step2}
    \begin{tabular}{lccc}
        \toprule
        \textbf{Class} & \textbf{Clean} & \textbf{Adversarial} & \textbf{Absolute Increase} \\
        \midrule
        \rowcolor{gray!30} \textbf{mushroom} & 18 & 19981 & \textbf{+39.93}\% \\
        echidna & 34 & 11675 & +23.28\% \\
        ocarina & 43 & 9361 & +18.64\% \\
        switch & 16 & 5232 & +10.43\% \\
        agaric & 68 & 952 & +1.77\% \\
        eggnog & 38 & 834 & +1.59\% \\
        chain mail & 43 & 456 & +0.83\% \\
        doormat & 51 & 330 & +0.56\% \\
        plastic bag & 32 & 207 & +0.35\% \\
        ashcan & 31 & 141 & +0.22\% \\
        monitor & 56 & 122 & +0.13\% \\
        crossword puzzle & 44 & 55 & +0.02\% \\
        space bar & 4 & 9 & +0.01\% \\
        file & 2 & 7 & +0.01\% \\
        cardigan & 0 & 0 & +0.00\% \\
        crane & 0 & 0 & +0.00\% \\
        maillot & 0 & 0 & +0.00\% \\
        brabancon griffon & 3 & 0 & $-$0.01\% \\
        ear & 3 & 0 & $-$0.01\% \\
        drake & 4 & 0 & $-$0.01\% \\
        \bottomrule
    \end{tabular}
\end{table}

\begin{table}[h]
\centering
\caption{Top 20 classes with the largest absolute increase under adversarial attack on Blended-poisoned model.}
\label{blended_step2}
\begin{tabular}{lccc}
\toprule
\textbf{Class} & \textbf{Clean} & \textbf{Adversarial} & \textbf{Absolute Increase} \\
\midrule
\rowcolor{gray!20}
\textbf{mushroom} & 18 & 30547 & \textbf{+61.06\%} \\
doormat & 64 & 2442 & +4.76\% \\
poncho & 56 & 1740 & +3.37\% \\
switch & 26 & 1564 & +3.08\% \\
eggnog & 37 & 1432 & +2.79\% \\
echidna & 39 & 570 & +1.06\% \\
ocarina & 41 & 558 & +1.03\% \\
sock & 78 & 494 & +0.83\% \\
pillow & 60 & 305 & +0.49\% \\
web site & 63 & 301 & +0.48\% \\
worm fence & 33 & 259 & +0.45\% \\
slug & 42 & 267 & +0.45\% \\
chain mail & 48 & 240 & +0.38\% \\
plunger & 37 & 190 & +0.31\% \\
mashed potato & 79 & 215 & +0.27\% \\
miniskirt & 74 & 180 & +0.21\% \\
hotdog & 49 & 152 & +0.21\% \\
cassette & 39 & 138 & +0.20\% \\
joystick & 40 & 132 & +0.18\% \\
monitor & 42 & 116 & +0.15\% \\
\bottomrule
\end{tabular}
\end{table}

\begin{table}[h]
    \centering
    \caption{Top 20 classes with the largest absolute increase under adversarial attack on SIG-poisoned model.}
    \label{sig_step2}
    \begin{tabular}{lccc}
        \toprule
        \textbf{Class} & \textbf{Clean} & \textbf{Adversarial} & \textbf{Absolute Increase} \\
        \midrule
        \rowcolor[gray]{0.9} \textbf{mushroom} & 76 & 18775 & \textbf{+37.40\%} \\
        agaric & 14 & 8808 & +17.59\% \\
        monitor & 54 & 5183 & +10.26\% \\
        modem & 65 & 1596 & +3.06\% \\
        chain mail & 48 & 1480 & +2.86\% \\
        thimble & 34 & 734 & +1.40\% \\
        airship & 40 & 653 & +1.23\% \\
        admiral & 18 & 551 & +1.07\% \\
        ocarina & 42 & 533 & +0.98\% \\
        echidna & 37 & 485 & +0.90\% \\
        joystick & 37 & 476 & +0.88\% \\
        desktop computer & 101 & 347 & +0.49\% \\
        file & 3 & 230 & +0.45\% \\
        ashcan & 37 & 242 & +0.41\% \\
        microwave & 59 & 249 & +0.38\% \\
        crate & 53 & 227 & +0.35\% \\
        switch & 16 & 171 & +0.31\% \\
        television & 71 & 220 & +0.30\% \\
        centipede & 27 & 155 & +0.26\% \\
        sidewinder & 25 & 141 & +0.23\% \\
        \bottomrule
    \end{tabular}
\end{table}

\begin{table}[h]
\centering
\caption{Top 20 classes with the largest absolute increase under adversarial attack on WaNet-poisoned model.}
\label{wanet_step2}
\begin{tabular}{lccc}
\toprule
\textbf{Class} & \textbf{Clean} & \textbf{Adversarial} & \textbf{Absolute Increase }\\
\midrule
\rowcolor{gray!20}
\textbf{mushroom} & 18 & 13206 & \textbf{+26.38\%} \\
agaric & 60 & 13056 & +25.99\% \\
chain mail & 48 & 5634 & +11.17\% \\
switch & 26 & 2008 & +3.96\% \\
echidna & 39 & 1578 & +3.08\% \\
admiral & 14 & 1059 & +2.09\% \\
joystick & 40 & 836 & +1.59\% \\
stinkhorn & 45 & 671 & +1.25\% \\
poncho & 56 & 568 & +1.02\% \\
file & 0 & 422 & +0.84\% \\
ocarina & 41 & 419 & +0.76\% \\
shovel & 53 & 407 & +0.71\% \\
projectile & 8 & 319 & +0.62\% \\
consomme & 67 & 372 & +0.61\% \\
eggnog & 37 & 317 & +0.56\% \\
carbonara & 60 & 325 & +0.53\% \\
armadillo & 46 & 256 & +0.42\% \\
Indian cobra & 28 & 222 & +0.39\% \\
microwave & 61 & 235 & +0.35\% \\
doormat & 64 & 234 & +0.34\% \\
\bottomrule
\end{tabular}
\end{table}

\begin{table}[h]
\centering
\caption{Top 20 classes with the largest absolute increase under adversarial attack on BadEncoder-poisoned model.}
\label{badencoder_step2}
\begin{tabular}{lccc}
\toprule
\textbf{Class} & \textbf{Clean} & \textbf{Adversarial} & \textbf{Absolute Increase} \\
\midrule
\rowcolor{gray!20}
\textbf{mushroom} & 422 & 2522 & \textbf{+4.20\%} \\
pillow & 94 & 384 & +0.58\% \\
mongoose & 91 & 374 & +0.57\% \\
toy poodle & 112 & 393 & +0.56\% \\
quail & 128 & 328 & +0.40\% \\
beagle & 103 & 292 & +0.38\% \\
poncho & 63 & 231 & +0.34\% \\
miniskirt & 90 & 254 & +0.33\% \\
Labrador retriever & 256 & 418 & +0.32\% \\
dingo & 126 & 281 & +0.31\% \\
slug & 40 & 187 & +0.29\% \\
desktop computer & 135 & 279 & +0.29\% \\
diaper & 105 & 246 & +0.28\% \\
amphibian & 18 & 152 & +0.27\% \\
rock python & 83 & 215 & +0.26\% \\
clog & 32 & 159 & +0.25\% \\
eel & 76 & 201 & +0.25\% \\
racer & 69 & 191 & +0.24\% \\
bloodhound & 54 & 175 & +0.24\% \\
mouse & 27 & 146 & +0.24\% \\
\bottomrule
\end{tabular}
\end{table}

\begin{table}[h]
    \centering
    \caption{Top 20 classes with the largest absolute increase under adversarial attack on BadCLIP-poisoned model.}
    \label{badclip_step2}
    \begin{tabular}{lccc}
        \toprule
        \textbf{Class} & \textbf{Clean} & \textbf{Adversarial} & \textbf{Absolute Increase} \\
        \midrule
        \rowcolor{gray!30} \textbf{mushroom} & 6 & 48623 & \textbf{+97.23\%} \\
        agaric & 67 & 1132 & +2.13\% \\
        maillot & 0 & 0 & +0.00\% \\
        crane & 0 & 0 & +0.00\% \\
        cardigan & 0 & 0 & +0.00\% \\
        space bar & 3 & 0 & $-$0.01\% \\
        ear & 3 & 0 & $-$0.01\% \\
        drake & 3 & 0 & $-$0.01\% \\
        brabancon griffon & 4 & 0 & $-$0.01\% \\
        horizontal bar & 4 & 0 & $-$0.01\% \\
        swab & 5 & 0 & $-$0.01\% \\
        black-footed ferret & 5 & 0 & $-$0.01\% \\
        file & 5 & 0 & $-$0.01\% \\
        lhasa & 6 & 0 & $-$0.01\% \\
        pickelhaube & 6 & 0 & $-$0.01\% \\
        nipple & 7 & 0 & $-$0.01\% \\
        mouse & 8 & 0 & $-$0.02\% \\
        appenzeller & 9 & 1 & $-$0.02\% \\
        bow & 8 & 0 & $-$0.02\% \\
        tiger cat & 8 & 0 & $-$0.02\% \\
        \bottomrule
    \end{tabular}
\end{table}

Based on the experimental results presented in Tables~\ref{badnet_step2}, \ref{blended_step2}, \ref{sig_step2} and \ref{badclip_step2}, we observe particularly pronounced adversarial effects in four attack scenarios. These effects, induced by adversarial perturbations, manifest as substantial shifts in the prediction frequency toward the target class. The BadCLIP attack induces the most significant shift, with a 97.23\% increase in mushroom prediction frequency, followed by Blended (61.06\%), BadNet (39.33\%), and SIG (37.40\%). As shown in Table~\ref{badencoder_step2}, even the relatively moderate BadEncoder attack (4.20\% increase) leads to a statistically significant bias, maintaining a 3.62 percentage point advantage over the second-most predicted class (“pillow” at 0.58\%).

Table~\ref{wanet_step2} reveals an intriguing pattern of taxonomic-specific vulnerability in the WaNet attack. The method produces nearly identical prediction increases for both the target ``mushroom'' category (26.38\%) and its taxonomically related counterpart ``agaric'' (25.99\%), with merely a 0.39 percentage point differential.
This remarkable similarity validates the relationship between the adversarial perturbations and the target class of the backdoor attack, as stated in the Main Text. From another perspective, the CLIP model's inherent semantic clustering enables the identification of a semantically similar class within the ImageNet-1K label space, even when the attacker's intended target does not fall within the 1,000 predefined categories, as discussed in Section~\ref{sec:label}. This property facilitates the subsequent steps of trigger inversion and activation tuning.

\subsection{Inverted Trigger Visualization Results for Six Attacks}

Our inversion results, shown in the bottom row of Figure~\ref{fig:comparison}, demonstrate two distinct spatial distribution patterns corresponding to different attack types. For localized trigger attacks (BadNet, BadCLIP, and BadEncoder), the inverted triggers maintain the characteristic bottom-right corner positioning observed in the original attacks. Conversely, for globally distributed attacks (Blended, SIG, and WaNet), the inverted triggers successfully reproduce the expected multi-region distribution patterns.

The spatial consistency between original and inverted triggers is evident in both cases, with the inverted versions achieving comparable attack success rates to their original counterparts. These results confirm that our inversion method preserves the essential spatial characteristics of different trigger types while maintaining their functional effectiveness.

\subsection{Key Layers Selected in Activation Tuning}

In this section, we present the layer selection results during the Activation Tuning process. Since different attacks show minimal variation in layer activation outcomes, we demonstrate the anomalous response layers for four distinct CLIP architectures (RN50, RN101, ViT-B/16, ViT-B/32) when confronted with inversion triggers, using the BadCLIP attack as the representative case.

The visual encoder of ResNet architectures consists of four residual layers. For these architectures, the analysis shows concentrated sensitivity in the final residual layers. As shown in Table~\ref{tab:RN50_layer_impact}, RN50 exhibits extreme sensitivity in \texttt{visual.layer4} with an impact value of 1.3802, which exceeds the significance threshold ($\mu + \sigma = 0.9914$) by 39.2\%. Similarly, Table~\ref{tab:RN_101layer_impact} reveals that RN101's \texttt{visual.layer4} shows comparable vulnerability with an impact of 1.1823, 38.5\% above its threshold of 0.8538. This final-layer concentration suggests that ResNet protections can focus on monitoring these critical bottlenecks.

\begin{table}[h!]
    \centering
    \caption{Layer impact analysis results for RN50.}
    \label{tab:RN50_layer_impact}
    \begin{tabular}{lrr}
        \toprule
        \textbf{Layer} & \textbf{Impact} & \textbf{Selected Key Layer} \\
        \midrule
        visual.layer1 & 0.1407 & No \\
        visual.layer2 & 0.1750 & No \\
        visual.layer3 & 0.1435 & No \\        \textbf{visual.layer4} & 1.3802 & \textbf{Yes} \\
        \midrule
        \multicolumn{2}{l}{\textbf{Significance Threshold}} & 0.9914 \\
        \multicolumn{2}{l}{\textbf{Mean}} & 0.4599 \\
        \multicolumn{2}{l}{\textbf{Std}} & 0.5315 \\
        \bottomrule
    \end{tabular}
\end{table}

\begin{table}[h!]
    \centering
    \caption{Layer impact analysis results for RN101.}
    \label{tab:RN_101layer_impact}
    \begin{tabular}{lrr}
        \toprule
        \textbf{Layer} & \textbf{Impact} & \textbf{Selected Key Layer} \\
        \midrule
        visual.layer1 & 0.1329 & No \\
        visual.layer2 & 0.1492 & No \\
        visual.layer3 & 0.1547 & No \\
        \textbf{visual.layer4} & 1.1823 & \textbf{Yes} \\
        \midrule
        \multicolumn{2}{l}{\textbf{Significance Threshold}} & 0.8538 \\
        \multicolumn{2}{l}{\textbf{Mean}} & 0.4048 \\
        \multicolumn{2}{l}{\textbf{Std}} & 0.4490 \\
        \bottomrule
    \end{tabular}
\end{table}

\begin{table}[h!]
    \centering
    \caption{Layer impact analysis results for ViT-B/16.}
    \label{tab:vit16_layer_impact}
    \begin{tabular}{lrr}
        \toprule
        \textbf{Layer} & \textbf{Impact} & \textbf{Selected Key Layer} \\
        \midrule
        visual.transformer.resblocks.0 & 0.0916 & No \\
        visual.transformer.resblocks.1 & 0.1458 & No \\
        visual.transformer.resblocks.2 & 0.2858 & No \\
        \textbf{visual.transformer.resblocks.3} & 0.3423 & \textbf{Yes} \\
        \textbf{visual.transformer.resblocks.4} & 0.3247 & \textbf{Yes} \\
        \textbf{visual.transformer.resblocks.5} & 0.2996 &\textbf{Yes} \\
        visual.transformer.resblocks.6 & 0.1895 & No \\
        visual.transformer.resblocks.7 & 0.1879 & No \\
        visual.transformer.resblocks.8 & 0.2002 & No \\
        visual.transformer.resblocks.9 & 0.1745 & No \\
        visual.transformer.resblocks.10 & 0.1959 & No \\
        visual.transformer.resblocks.11 & 0.1700 & No \\
        \midrule
        \multicolumn{2}{l}{\textbf{Significance Threshold}} & 0.2915 \\
        \multicolumn{2}{l}{\textbf{Mean}} & 0.2173 \\
        \multicolumn{2}{l}{\textbf{Std}} & 0.0742 \\
        \bottomrule
    \end{tabular}
\end{table}

\begin{table}[h!]
    \centering
    \caption{Layer impact analysis results for ViT-B/32.}
    \label{tab:vit32_layer_impact}
    \begin{tabular}{lrr}
        \toprule
        \textbf{Layer} & \textbf{Impact} & \textbf{Selected Key Layer} \\
        \midrule
        visual.transformer.resblocks.0 & 0.0965 & No \\
        visual.transformer.resblocks.1 & 0.2025 & No \\
        visual.transformer.resblocks.2 & 0.3066 & No \\
        \textbf{visual.transformer.resblocks.3} & 0.3782 & \textbf{Yes} \\
       \textbf{visual.transformer.resblocks.4} & 0.3609 & \textbf{Yes} \\
        visual.transformer.resblocks.5 & 0.3085 & No \\
        visual.transformer.resblocks.6 & 0.2558 & No \\
        visual.transformer.resblocks.7 & 0.2628 & No \\
        visual.transformer.resblocks.8 & 0.2463 & No \\
        visual.transformer.resblocks.9 & 0.2172 & No \\
        visual.transformer.resblocks.10 & 0.2334 & No \\
        visual.transformer.resblocks.11 & 0.1367 & No \\
        \midrule
        \multicolumn{2}{l}{\textbf{Significance Threshold}} & 0.3298 \\
        \multicolumn{2}{l}{\textbf{Mean}} & 0.2504 \\
        \multicolumn{2}{l}{\textbf{Std}} & 0.0794 \\
        \bottomrule
    \end{tabular}
\end{table}

The CLIP visual encoder using the ViT-B architecture consists of 12 Transformer blocks, from which we identify key layers for analysis. Transformer architectures display fundamentally different response patterns characterized by distributed sensitivity across middle layers. In ViT-B/16 (Table~\ref{tab:vit16_layer_impact}), blocks 3–5 show consistent anomalous responses (0.3423, 0.3247, 0.2996) that all exceed the threshold of 0.2915. The ViT-B/32 architecture (Table~\ref{tab:vit32_layer_impact}) reveals similar distributed sensitivity, with blocks 3–4 showing the strongest deviations (0.3782, 0.3609), surpassing the threshold of 0.3298 by 14.7\% and 9.4\%, respectively. This pattern correlates with the attention mechanism’s global dependency formation in intermediate layers, requiring defense strategies that monitor multiple blocks rather than single points of failure.

The statistical robustness of our $\mu + \sigma$ selection criterion is confirmed by consistent performance across all architectures, with all identified key layers showing notable deviations. The clear separation between normal and anomalous layers (minimum margin of 9.4\%) demonstrates the method's reliability for architecture-agnostic backdoor analysis. These findings suggest that effective defense strategies must account for fundamental architectural differences—implementing focused final-layer monitoring for ResNets versus comprehensive multi-block analysis for Transformers.

\section{Limitations}
\label{sec6}

While our study offers valuable insights, there are certain limitations that could be addressed in future research:

First, our analysis primarily focuses on the CLIP framework, examining various CLIP architectures such as RN50, RN101, ViT-B/16, and ViT-B/32. While these models are representative of the CLIP family, our work does not explore other large-scale multimodal learning architectures, which may exhibit different characteristics in terms of vulnerabilities or defense strategies. This focus on CLIP models leaves open the potential for discovering broader patterns across other architectures in future studies.

Second, our approach to adversarial sample generation relies on methods proposed by AdvCLIP~\cite{zhou2023advclip}. While this provides a solid foundation for our analysis, the range of adversarial generation techniques employed could be expanded. Exploring alternative attack methodologies may offer a more comprehensive understanding of how different adversarial strategies interact with multimodal models, enriching the robustness of our findings.

These areas of future work suggest opportunities to broaden the scope of the research, providing a more holistic view of both multimodal architectures and adversarial generation methods.


\end{document}